\documentclass[12pt,epsf]{article}
\usepackage{amsmath,amssymb}
\usepackage{graphicx}
\usepackage{cite,fancyhdr}
\usepackage{braket}
\usepackage{url}
\usepackage{cancel}
\usepackage{color}
\usepackage{tcolorbox}
\definecolor{light-gray}{gray}{1}
\usepackage{mathtools} 
\usepackage[
colorlinks=true, 
linkcolor=RoyalBlue, 
citecolor=RoyalBlue,
urlcolor=black]{hyperref}
\usepackage[left=3cm,right=3cm,top=3.5cm,bottom=3.5cm,a4paper]{geometry}

\definecolor{Orange}{cmyk}{0,0.61,0.87,0}
\definecolor{JungleGreen}{cmyk}{0.99,0,0.52,0}
\definecolor{OliveGreen}{cmyk}{0.64,0,0.95,0.40}
\definecolor{Brown}{cmyk}{0,0.81,1,0.60}
\definecolor{RoyalBlue}{cmyk}{0.71,0.53,0,0.12}
\definecolor{Gray}{cmyk}{0,0,0,0.40}
\definecolor{Pink}{cmyk}{0.0,1,0,0}
\definecolor{LightPink}{cmyk}{0.0,0.25,0,0}
\definecolor{LLightPink}{cmyk}{0.0,0.10,0,0}
\definecolor{LightBlue}{cmyk}{0.25,0,0,0}
\definecolor{LightGray}{cmyk}{0,0,0,0.2}
\definecolor{LightGreen}{cmyk}{0.3,0,0.3,0}
\def\l{\left}
\def\r{\right}
\def\nn{\nonumber}

\def\refheft{\cite{
Feruglio:1992wf,Burgess:1999ha,Giudice:2007fh,Grinstein:2007iv,Alonso:2012px,Buchalla:2012qq,Azatov:2012bz,Contino:2013kra,Jenkins:2013fya,Buchalla:2013rka,Alonso:2014rga,Guo:2015isa,Buchalla:2015qju,Alonso:2017tdy,Buchalla:2017jlu,Buchalla:2018yce}}
\def\refsmeft{\cite{Buchmuller:1985jz, Grzadkowski:2010es, Brivio:2017vri, Manohar:2018aog}}
\allowdisplaybreaks
\begin{document}
\setcounter{footnote}{0}
\setcounter{figure}{0}
\setcounter{table}{0}
\renewcommand{\thefootnote}{\fnsymbol{footnote}}
\numberwithin{equation}{section}
\begin{titlepage}
\begin{flushright}
{\tt 
OU-HET-1122
}
\end{flushright}
\vskip 3.35cm
\begin{center}
{
\large{\bf
A new Higgs effective field theory and \\
the new no-lose theorem
}
}
\vskip 1.2cm
Shinya Kanemura{\footnote{
E-mail address:~{\tt{kanemu@het.phys.sci.osaka-u.ac.jp}}}}
and
Ryo Nagai{\footnote{
E-mail address:~{\tt{nagai@het.phys.sci.osaka-u.ac.jp}}}}
\vskip 0.4cm
{\small \it
Department of Physics, Osaka University, Toyonaka, Osaka 560-0043, Japan
}\\ [3pt] 
(\today)
\vskip 1.5cm
\begin{abstract}
Non-decoupling effects of heavy new particles cannot be described by the standard effective field theory with finite truncation of higher dimensional operators.
We propose a new effective field theory 
in which non-decoupling quantum effects of new physics are 
correctly described.
We discuss vacuum stability and perturbative unitarity in our effective field theory, 
and we find that
the scale of new physics can be estimated if we will observe the Higgs coupling deviation via non-decoupling effects in future collider experiments.
\end{abstract}
\end{center}
\end{titlepage}
\renewcommand{\thefootnote}{\roman{footnote}}
\setcounter{footnote}{0}
\section{Introduction}
One of the big mysteries in particle physics is the origin of electroweak symmetry breaking (EWSB). 
In the standard model (SM) of particle physics, 
the EWSB is just assumed to occur due to a given scalar potential. 
This artificial aspect causes the hierarchy problem between the electroweak scale and the fundamental ultraviolet (UV) scale such as the Plank or Grand Unification scales.
In addition, there are several important phenomena which cannot be explained in the SM, such as neutrino oscillation, dark matter,
baryon asymmetry of the Universe and so on. 
We therefore believe that the SM should be extended.

However, we have not seen direct signals of physics beyond the SM yet.
The direct searches of new particles at the high-energy collider experiments have set severe constraints on the mass scale of new particles, which roughly reach beyond the TeV scales.
This fact implies that the mass scale of the new particle is much above the electroweak scale.

Does non-observation of new particles at the direct searches indicate the nightmare scenario,
where we cannot observe any signal of the new physics at future experiments? 
The answer must be {\it{no}}.
One may answer ``{\it{yes}}''  based on the conventional decoupling picture 
that all effects of the heavy new particle to the low energy observables are suppressed by powers of the mass of the new particle \cite{Appelquist:1974tg}. 
However, the decoupling theorem \cite{Appelquist:1974tg} does not work if the heavy particles obtain their masses predominantly from the interactions with Higgs field. 
The new particle effects to the low-energy observables, in this case, are not suppressed in the heavy limit, rather it can be enhanced by power-like contributions of their masses. 
This kind of new physics effects is categorized as ``non-decoupling effects''.
We will be concerned with non-decoupling effects of new physics beyond the SM in this paper.

The non-decoupling effects have been gathered a lot of attentions as important signatures of undiscovered heavy particles in low-energy observables. 
For example, before the discoveries of the top quark and the Higgs boson, 
we could estimate their mass scales from the precise measurements of 
properties of the electroweak gauge bosons.  
The electroweak oblique corrections are sensitive to the non-negative power of 
the masses of the top quark and the Higgs boson due to their non-decoupling effects \cite{Kennedy:1988sn, Peskin:1990zt, Altarelli:1990zd, Peskin:1991sw, Altarelli:1991fk}, 
by which their masses were approximately known before their direct descoveries using the precision measurement at LEP, SLC and so on. 
We could also study the possibility of the new heavy particles by focusing on 
the non-decoupling effects in the electroweak gauge sector. 
See Refs.~\cite{
Inami:1980fz,
Peskin:1991sw,
Inami:1994nj,
Inami:1995ep
} for examples.

The non-decoupling effects also appear in the low-energy observables relating with the 125\,GeV Higgs boson,
such as $hVV$, $hff$, and $hhh$ coupling strengths with $h$, $V$, and $f$ being the 125\,GeV Higgs boson, gauge bosons, and fermions respectively. 
For example, in Refs.~\cite{
Kanemura:1997ej,
Kanemura:2002vm,
Kanemura:2004mg,
Aoki:2012jj,
Kanemura:2014dja,
Kanemura:2015fra,
Kanemura:2016lkz,
Kanemura:2018yai,
Braathen:2019pxr,
Braathen:2019zoh
},
the non-decoupling effects in concrete extended Higgs models were estimated in detail,
and it was shown that the new heavy particles effects appear as the positive power of their heavy masses 
if the new particles obtain mainly from the Higgs VEV. 
These non-decoupling effects should be important signitures of the heavy particles at
the precise measurements of the Higgs couplings, 
which are on-going at the collider experiments and expected to be more precisely performed at the future colliders.

How can we treat the theoretical predictions of non-decoupling effects in a model-independent way? 
Considering the scale hierarchy between the electroweak scale and the masses of new particles, 
it is natural to employ Effective Field Theory (EFT) framework consisted of the SM fields to parameterize the new physics effects.
One of the widely used EFT frameworks is the ``Standard Model Effective Field Theory (SMEFT) \refsmeft'', in which new physics effects are expressed in polynomials of the SM fields transforming linearly under the electroweak gauge symmetry ($SU(2)_L\times U(1)_Y$) transformation. 
It has been recently pointed out, however, that the SMEFT cannot be applied to describe the non-decoupling physics \cite{Alonso:2016oah, Falkowski:2019tft,Cohen:2020xca,Cohen:2021ucp}. 
This is because the non-decoupling effects in the Higgs sector typically cannot be expressed by a polynomial in the Higgs doublet fields. In order to describe non-decoupling effects due to new physics in the Higgs sector, we have to employ a more general Higgs EFT framework. 

In this paper we propose a new EFT framework, where non-decoupling new physics effects in the Higgs sector can be systhematically described.
We employ the effective potential formalism by Coleman and Weinberg \cite{Coleman:1973jx} to parameterize non-decoupling quantum effects in the Higgs potential.
We also describe the non-decoupling effects in the parts other than the Higgs potential by 
using polynomials in the 125\,GeV Higgs field instead of $SU(2)_L$ doublet Higgs field. 
Our new effective theory is an extension of so-called the ``Higgs effective field theory (HEFT)'' \refheft~
such that the Higgs potential is extended so as to contain quantum effects of new physics.

Using our new EFT framework, we discuss the relationship between 
the new physics scale and the size of the Higgs coupling deviations caused by the non-decoupling effect.
We consider the two theoretical constraints from vacuum stability and perturbative unitarity.
Regarding the perturbative unitarity, we calculate $S$-wave amplitudes for two-body elastic scatterings of Nambu-Goldstone (NG) bosons and physical scalar bosons at high energies, and impose the perturbative unitarity bound on the maximum eigenvalue of the scattering matrix. Combining the vacuum stability and perturbative unitarity bounds, we obtain the upper bound on the scale of new physics as a function of the Higgs coupling deviations.

Our finding means that once we observe the Higgs coupling deviation at future collider experiments 
we can estimate the upper bound of the new physics scale which should be a target energy scale for future colliders designed to observe the new particles directly.
Our argument is an extension of the classic argument by Lee, Quigg, and Thacker in 1977 \cite{Lee:1977eg,Lee:1977yc}. 
They predicted the existence of new physics below the TeV scale in the theory without a Higgs sector by imposing perturbative unitarity in the scatterings of longitudinally polarized $W$ and $Z$ bosons.
Their arguments gave one of the most important motivations for exploring the energy scale of the LHC, which was refereed as the ``no-lose theorem''. 
We extend their argument by including non-decoupling effects to the Higgs boson couplings. 
Inspired by their successful argument, we refer to our findings mentioned above as the ``{\it{new}} no-lose theorem'', which can be applied to exploring the Higgs sector if we observe the Higgs coupling deviation at future collider experiments.

This paper is organized as follows. 
We define our new EFT Lagrangian
and estimate the Higgs coupling deviation factors in section \ref{sec:naHEFT}. 
In section \ref{sec:model-scalar}, we introduce a benchmark UV completion of our EFT and compare the EFT results with the full model calculation. We then confirm that naHEFT successfully describes the non-decoupling effects which cannot be parameterized by SMEFT approximation.
We discuss vacuum stability and perturbative unitarity in section \ref{sec:vacuum stability} and \ref{sec:unitarity}, respectively. 
Numerical results for the vacuum stability and perturbative unitarity bounds are shown in section \ref{sec:numeric}. 
We summarize our findings and propose the ``new no-lose theorem'' for future colliders in section \ref{sec:nolose}.
Finally, we give our conclusion with mentioning the outlook in section \ref{sec:summary}.

\section{Nearly aligned Higgs EFT}
\label{sec:naHEFT}
We introduce an effective theory 
in which non-decoupling new physics effects are 
correctly described.
We assume that the new physics scale is higher than the electroweak scale, and that the particle content is the same as the SM one.
Our new effective theory is an extension of so-called the ``Higgs effective field theory'' \refheft~such that the Higgs potential is extended to contain quantum effects of new physics.
Corresponding to the current experimental data from ATLAS \cite{ATLAS:2019nkf} and CMS \cite{CMS:2020gsy}, 
which indicate that the measured Higgs coupling constants are SM like,
we assume that the new physics effects enter into the low-energy theory via the quantum effects.
In this case, 
deviations from the SM in Higgs coupling constants
with gauge bosons and fermions appear in the loop corrections.
We here call the effective theory describing this scenario as the ``nearly aligned Higgs effective field theory (naHEFT)''.

The effective Lagrangian is given as 
\begin{align}
\mathcal{L}_{\rm{naHEFT}}
&\,=\,
\mathcal{L}_{\rm{SM}}
\,+\,
\mathcal{L}_{\rm{BSM}}
\,,
\label{eq:naHEFT}
\end{align}
where $\mathcal{L}_{\rm{SM}}$ is the Lagrangian of the SM, 
and $\mathcal{L}_{\rm{BSM}}$ is defined by
\begin{align}
\mathcal{L}_{\rm{BSM}}
&\,=\,
\xi\,\biggl[
-
\frac{\kappa_0}{4}
\,[\mathcal{M}^2(h)]^2\,
\ln\frac{\mathcal{M}^2(h)}{\mu^2}
\nn\\
&\qquad
+
\frac{v^2}{2}
\,\mathcal{F}(h)\,
\mbox{Tr}[D_\mu U^\dag D^\mu U]
+
\frac{1}{2}
\,\mathcal{K}(h)
(\partial_\mu h)(\partial^\mu h)
\nn\\
&\qquad
\,-\,
v
\biggl(\bar{q}^i_L  U \! \l[\mathcal{Y}^{ij}_{q}(h)+\hat{\mathcal{Y}}^{ij}_q(h)\tau^3\r] q^j_R+h.c.\biggr)
\nn\\
&\qquad
\,-\,
v
\biggl(\bar{l}^i_L  U \! \l[\mathcal{Y}^{ij}_{l}(h)+\hat{\mathcal{Y}}^{ij}_l(h)\tau^3\r] l^j_R+h.c.\biggr)
\biggr]
\,,
\label{eq:BSM}
\end{align}
with $\xi=1/(4\pi)^2$.
$\kappa_0$ and $\mu^2$ are real parameters.
We take $v\simeq 246\,$GeV.
$h$ denotes the 125\,GeV Higgs boson, and
we here assume $h=0$ to be the global minimum of the Higgs potential.
We will discuss the validity of this assumption later.
$U$ parameterizes the Nambu-Goldstone (NG) bosons $(\pi^\pm,\pi^3)$ eaten by the longitudinal $W^\pm$ and $Z$ bosons,
\begin{align}
U\,=\,
\exp\l(\frac{i}{v}\pi^a \tau^a\r)
\,,\qquad
\pi^\pm
\,=\,
\frac{1}{\sqrt{2}}(\pi^1\mp i\pi^2)
\,,
\end{align}
with $\tau^a~(a=1,2,3)$ being the $SU(2)$ Pauli matrices.
$q^i_L$ and $l^i_L$ denote the $SU(2)_L$ doublet SM quark and lepton fields, respectively. 
$i$ is the index for the generation, $i=1,2,3$. 
$q^i_R$ and $l^i_R$ are vectors defined as 
$q^i_R=(u^i_R~d^i_R)^T$ and $l^i_R=(0~e^i_R)^T$ where
$u^i_R$, $d^i_R$, and $e^i_R$ are
the $SU(2)_L$ singlet up-type quark, down-type quark and lepton fields, respectively.
The covariant derivative of $U$ is defined as
\begin{align}
D_\mu U
\,=\,
\partial_\mu U
+ig{\bf{W}}_\mu U
-ig' U {\bf{B}}_\mu 
\,,
\end{align}
where $SU(2)_L$ and $U(1)_Y$ gauge boson fields are defined as
${\bf{W}}_\mu=\sum_{a=1}^3 W^a_\mu \frac{\tau^a}{2}$ and
${\bf{B}}_\mu=B_\mu \frac{\tau^3}{2}$. 
$g$ and $g'$ denote the $SU(2)_L$ and $U(1)_Y$ gauge couplings. 
$\mathcal{M}^2(h)$, $\mathcal{F}(h)$, $\mathcal{K}(h)$, $\mathcal{Y}(h)$, and $\hat{\mathcal{Y}}(h)$
 are polynomial in $h$,
\begin{align}
&
\mathcal{M}^2(h)
\,=\,
\sum_{n=0}\,
\frac{\mathcal{M}_n}{n!}
\l(\frac{h}{v}\r)^n
\,,\\
&
\mathcal{F}(h)
\,=\,
\sum_{n=1} \,
\frac{f_n}{n!}
\l(\frac{h}{v}\r)^n
\,,\quad
\mathcal{K}(h)
\,=\,
\sum_{n=0} \,
\frac{k_n}{n!}
\l(\frac{h}{v}\r)^n
\,,\\
&
\mathcal{Y}^{ij}_\psi(h)
\,=\,
\sum_{n=1} \,
\frac{y^{ij}_{\psi,n}}{n!}
\l(\frac{h}{v}\r)^n
\,,\quad
\hat{\mathcal{Y}}^{ij}_\psi(h)
\,=\,
\sum_{n=1} \,
\frac{{\hat{y}}^{ij}_{\psi,n}}{n!}
\l(\frac{h}{v}\r)^n
\,,
\end{align}
where $\psi=q,l$. 
The coefficients $\mathcal{M}_n$, $f_n$ and $k_n$ are real, 
while $y^{ij}_{\psi,n}$ and $\hat{y}^{ij}_{\psi,n}$ can be complex.

Under the $G_{\rm{EW}}=SU(2)_L\times U(1)_Y$ gauge transformation, the SM fields transform as
\begin{align}
&h
\xrightarrow{G_{\rm{EW}}}
h\,,\qquad
U
\xrightarrow{G_{\rm{EW}}}
 \mathfrak{g}_L U \mathfrak{g}^\dag_Y\,,\qquad
\\
&
{\bf{W}}_\mu
\xrightarrow{G_{\rm{EW}}}
\mathfrak{g}_L {\bf{W}}_\mu \mathfrak{g}^\dag_L
+
\frac{i}{g}(\partial_\mu \mathfrak{g}_L) \mathfrak{g}^\dag_L
\,,
\\
&
{\bf{B}}_\mu
\xrightarrow{G_{\rm{EW}}}
\mathfrak{g}_Y {\bf{B}}_\mu \mathfrak{g}^\dag_Y
+
\frac{i}{g'}(\partial_\mu \mathfrak{g}_Y) \mathfrak{g}^\dag_Y
\,,
\\
&q^i_L
\xrightarrow{G_{\rm{EW}}}
 e^{\frac{i}{6}\theta_Y}\mathfrak{g}_L q^i_L\,,\qquad
q^i_R
\xrightarrow{G_{\rm{EW}}}
 e^{\frac{i}{6}\theta_Y}\mathfrak{g}_Y q^i_R\,,\qquad
\\
&l^i_L
\xrightarrow{G_{\rm{EW}}}
 e^{-\frac{i}{2}\theta_Y}\mathfrak{g}_L l^i_L\,,\qquad
l^i_R
\xrightarrow{G_{\rm{EW}}}
 e^{-\frac{i}{2}\theta_Y}\mathfrak{g}_Y l^i_R\,,\qquad
\end{align}
where
\begin{align}
\mathfrak{g}_L=\exp\l(i\theta^a\frac{\tau^a}{2}\r)\,,\qquad
\mathfrak{g}_Y=\exp\l(i\theta_Y\frac{\tau^3}{2}\r)\,,
\end{align}
with $\theta^a_L$ and $\theta_Y$ being real parameters.
We note that the Lagrangian (\ref{eq:naHEFT}) is invariant under the $G_{\rm{EW}}=SU(2)_L\times U(1)_Y$ gauge transformation.

We put the one-loop suppression factor $\xi=1/(4\pi)^2$ in the BSM part 
to reflect our interested scenario where the dominant new physics effects appear at one loop-level. To keep the consistency with the loop expansion, 
we regard $\xi$ as an expansion parameter in the EFT analysis.
Hereafter, we neglect the $\mathcal{O}(\xi^2)$ corrections. 

New physics effects are encoded in polynomials, $\mathcal{M}^2(h)$, $\mathcal{F}(h)$, $\mathcal{K}(h)$, $\mathcal{Y}(h)$, and $\hat{\mathcal{Y}}(h)$. 
If there is no non-decoupling effects, 
the effective Lagrangian is expressed by polynomials in $|\Phi|^2=(v+h)^2/2$ rather than polynomials in $h$ \cite{Alonso:2016oah, Falkowski:2019tft,Cohen:2020xca,Cohen:2021ucp}. 
The EFT described by polynomials in $|\Phi|^2$ is known as the ``Standard Model Effective Field Theory (SMEFT) \refsmeft''.

For the Higgs potential part in $\mathcal{L}_{\rm{BSM}}$, 
we parameterize the BSM corrections by 
the famous Coleman-Weinberg potential \cite{Coleman:1973jx}.
We employ their formalism to parameterize non-decoupling quantum effects of 
new physics in a systematic way. 
$\mathcal{M}^2(h)$ encodes the information how the new particles obtain their masses. Here we assume that the masses of the new particles are universal just for simplicity.
Different new physics scale can be easily introduced by adding the same term with the different polynomial in the BSM part.
We define the scale $\Lambda$ as
\begin{align}
\Lambda^2
\simeq
\mathcal{M}^2(0)=\mathcal{M}_0\,,
\label{eq:defLam}
\end{align}
which
corresponds to the mass squared of the integrated new particles. 
We assume $\mathcal{M}^2(0)>0$.
We regard $\Lambda$ as the cutoff scale of our EFT.
We therefore assume 
\begin{align}
\Lambda
 > v
\,,
\label{eq:cutoff}
\end{align}
to ensure the validity of the EFT description.
$\kappa_0$ corresponds to the effective degree of freedom of the new particles
contributing to the Higgs potential.
Positive and negative values of $\kappa_0$ imply bosonic and fermionic contributions, respectively.
$\mu$ can be regarded as the renormalization scale for the one-loop corrections to the Higgs potential. 
The physical observables should not depend on the scale $\mu$ after performing on-shell renormalization. We will discuss this point later.

$\mathcal{F}(h)$ modifies the Higgs couplings to the electroweak gauge bosons and NG bosons. 
We take $\mathcal{F}(0)=0$ to make the kinetic terms of the NG bosons be canonical ones. 
Here we assume that the Higgs-gauge sector respects a global $SU(2)$ custodial symmetry except for $U(1)_Y$ gauge interaction. If the BSM Higgs-gauge sector breaks the custodial symmetry explicitly, $\mathcal{F}_Z(h)\,(\mbox{Tr}[U^\dag D_\mu U \tau^3])^2$ with $\mathcal{F}_Z(h)$ being a polynomial in $h$ appears.

$\mathcal{Y}(h)$ and $\hat{\mathcal{Y}}(h)$ result in the Higgs-fermion interactions which are not predicted in the SM.
Here we take $\mathcal{Y}(0)=\hat{\mathcal{Y}}(0)=0$ so that the fermion mass parameters are fixed to be ones in the SM.

$\mathcal{K}(h)$ expresses the new physics effects to the wave function of $h$. We assume
\begin{align}
1+\xi\,k_0>0
\,.
\end{align}
to avoid the negative kinetic energy of $h$.
The nonzero $\mathcal{K}(h)$ induces the universal deviation of Higgs coupling constants via canonical normalization of the Higgs field. 
The canonically normalized Higgs field ${\hat{h}}$ 
is given as
\begin{align}
\hat{h}
&\,=\,
\int^h_0 dh'\sqrt{1+\xi\,\mathcal{K}(h')}
\,.
\label{eq:hhat}
\end{align}
We find
\begin{align}
h\,=\,\sum_{n=1}^\infty \frac{{c}_n}{n!}\hat{h}^n
\,,
\label{eq:htohhat}
\end{align}
with 
\begin{align}
{c}_1
&\,=\,
1-\frac{\xi}{2}k_0
+
\mathcal{O}(\xi^2)
\,,
\quad
{c}_n
\,=\,
-\frac{\xi}{2}\frac{k_{n-1}}{v^{n-1}}
+
\mathcal{O}(\xi^2)
\quad 
\mbox{for}~n\geq 2
\,.
\end{align}

In the naHEFT, 
$n$-point functions of Higgs field at the zero momentum are obtained by 
$-\frac{\partial^n {\mathcal{L}_{\rm{naHEFT}}}}{\partial \hat{h}^n}|_{\hat{h}=0}$. 
Here we assume 
\begin{align}
-\frac{\partial{\mathcal{L}_{\rm{naHEFT}}}}{\partial \hat{h}}\biggl|_{\hat{h}=0}
\,=\,
c_1 d_1
\,=\,
0
\,,\label{eq:dhat1}
\end{align}
where 
$d_n$ is defined as
\begin{align}
d_n=-\frac{\partial^n \mathcal{L}_{\rm{naHEFT}}}{\partial^n h}\biggl|_{h=0}
\,.
\end{align}
The Higgs mass squared $M^2_h$ is given as
\begin{align}
M^2_h
\,=\,
-\frac{\partial^2 {\mathcal{L}_{\rm{naHEFT}}}}{\partial \hat{h}^2}\biggl|_{\hat{h}=0}
\,=\,
c^2_1 d_2
\,.
\label{eq:Mhsq}
\end{align}
The positivity of $M^2_h$ is ensured by the vacuum condition.
The $n\geq 3$ point Higgs coupling parameters are obtained as
\begin{align}
-\frac{\partial^3 {\mathcal{L}_{\rm{naHEFT}}}}{\partial \hat{h}^3}\biggl|_{\hat{h}=0}
&\,=\,
c^3_1 d_3 
+ 
3c_1c_2 d_2
\,,\\
-\frac{\partial^4 {\mathcal{L}_{\rm{naHEFT}}}}{\partial \hat{h}^4}\biggl|_{\hat{h}=0}
&\,=\,
c^4_1 d_4
+
6c^2_1 c_2 d_3
+
4c_1 c_3 d_2
+
3c^2_2 d_2
\,,\\
&
~\vdots
\nn
\end{align} 
where we used Eq.~(\ref{eq:dhat1}).

In the canonical basis, the naHEFT Lagrangian is expanded as
\begin{align}
\mathcal{L}_{\rm{naHEFT}}
\,=\,
&
-\frac{1}{4}W^{a\mu\nu}W^a_{\mu\nu}
-\frac{1}{4}B^{\mu\nu}B_{\mu\nu}
\nn\\
&
+
\frac{v^2}{4}
\l(1
+2\kappa_V \frac{\hat{h}}{v}
+\kappa_{VV} \frac{\hat{h}^2}{v^2}
+\mathcal{O}(\hat{h}^3)
\r)\,
\mbox{Tr}
[D_\mu U^\dag D^\mu U]
\nn\\
&
\,+
\frac{1}{2}(\partial_\mu \hat{h})(\partial^\mu \hat{h})
-
\frac{1}{2}M^2_h \hat{h}^2
-
\frac{1}{3!}\frac{3M^2_h}{v}\kappa_3 \hat{h}^3
-
\frac{1}{4!}\frac{3M^2_h}{v^2}\kappa_4 \hat{h}^4
+\mathcal{O}({h}^5)
\nn\\
&
-
\sum_{f=u,d,e}m_{f^i} 
\biggl[
\l(\delta^{ij}+\kappa^{ij}_f\frac{h}{v}+\mathcal{O}(h^2,\pi^2)\r)\bar{f}^i_L f^j_R+h.c.
\biggr]
\,,
\label{eq:Lint}
\end{align}
where $m_{f^i}$ denotes the positive real mass parameter for the SM fermion, $f^i$.
The $\kappa$ parameters are obtained as
\begin{align}
\kappa_{V}
&\,=\,
c_1\l(1+\frac{\xi}{2}f_1\r)
\,,\\
\kappa_{VV}
&\,=\,
c^2_1\l(1+\frac{\xi}{2}f_2\r)
+c_2 v\l(1+\frac{\xi}{2}f_1\r)
\,,\\
\kappa^{ij}_u
&\,=\,
c_1\biggl(\delta^{ij}
+\xi\,
(y^{ij}_{q,1}
+y^{ij}_{\bar{q},1})
\frac{
v
}{m_{u^i}}
\biggr)
\,,\\
\kappa^{ij}_d
&\,=\,
c_1\biggl(\delta^{ij}
+\xi\,
(y^{ij}_{q,1}
-y^{ij}_{\bar{q},1})
\frac{
v
}{m_{d^i}}
\biggr)
\,,\\
\kappa^{ij}_e
&\,=\,
c_1\biggl(\delta^{ij}
+\xi\,
(y^{ij}_{l,1}
-y^{ij}_{\bar{l},1})
\frac{
v
}
{m_{e^i}}
\biggr)
\,,\\
\kappa_{3}
&\,=\,
\frac{v}{3M^2_h}
\l(
c^3_1 d_3 
+ 
3c_1c_2 d_2
\r)
\,,\\
\kappa_4
&\,=\,
\frac{v^2}{3M^2_h}\l(
c^4_1 d_4
+
6c^2_1 c_2 d_3
+
4c_1 c_3 d_2
+
3c^2_2 d_2
\r)
\,.
\end{align}
Ignoring the $\mathcal{O}(\xi^2)$ corrections, we find
\begin{align}
\kappa_{V}
&\,=\,
1+\Delta\kappa_V
\,,\qquad
\kappa_{VV}
\,=\,
1+\Delta\kappa_{VV}
\,,\\
\kappa^{ij}_{f}
&\,=\,
\delta^{ij}+\Delta\kappa^{ij}_{f}
\,,\qquad f=u,d,e
\,\\
\kappa_{3}
&\,=\,
1+\Delta\kappa_{3}
\,,\qquad
\kappa_{4}
\,=\,
1+\Delta\kappa_{4}
\,,
\end{align}
where
\begin{align}
\Delta\kappa_V
&\,=\,
\frac{\xi}{2}( f_1-k_0)
\,,
\label{eq:kappaV-naHEFT}
\\
\Delta\kappa_{VV}
&\,=\,
\frac{\xi}{2} \l(
f_2-k_1-2k_0
\r)
\,,
\label{eq:kappaVV-naHEFT}
\\
\Delta\kappa^{ij}_{u}
&\,=\,
\xi
\l[
(y^{ij}_{q,1}+y^{ij}_{\bar{q},1})
\frac{
v
}{m_{u^i}}
-
\frac{k_0}{2}\delta^{ij}
\r]
\,,\\
\Delta\kappa^{ij}_{d}
&\,=\,
\xi
\l[
(y^{ij}_{q,1}-y^{ij}_{\bar{q},1})
\frac{
v
}{m_{d^i}}
-
\frac{k_0}{2}\delta^{ij}
\r]
\,,\\
\Delta\kappa^{ij}_{e}
&\,=\,
\xi
\l[
(y^{ij}_{l,1}-y^{ij}_{\bar{l},1})
\frac{
v
}{m_{l^i}}
-
\frac{k_0}{2}\delta^{ij}
\r]
\,,\\
\Delta\kappa_{3}
&\,=\,
-\frac{\xi}{2} (k_0+k_1)
\nn\\
&
\quad
+
\frac{\xi\kappa_0}{6v^2M^2_h}
\biggl[
3(\mathcal{M}_0-\mathcal{M}_1)(\mathcal{M}_1-\mathcal{M}_2)
+\mathcal{M}_0\mathcal{M}_3
\biggr]
\ln\frac{\mathcal{M}_0}{\mu^2}
\nn\\
&
\quad
+
\frac{\xi\kappa_0}{12v^2M^2_h}
\biggl[
2\frac{\mathcal{M}^3_1}{\mathcal{M}_0}
-9\mathcal{M}_1(\mathcal{M}_1-\mathcal{M}_2)
+\mathcal{M}_0(3\mathcal{M}_1-3\mathcal{M}_2+\mathcal{M}_3)
\biggr]
\,,\\
\Delta\kappa_4
&\,=\,
-\xi \l(k_0+\frac{2}{3}k_2\r)
\nn\\
&
\quad
+
\frac{\xi \kappa_0}{6v^2M^2_h}
\biggl[
3(\mathcal{M}^2_2-\mathcal{M}^2_1)
+4\mathcal{M}_1\mathcal{M}_3
+\mathcal{M}_0(3\mathcal{M}_1-3\mathcal{M}_2+\mathcal{M}_4)
\biggr]
\ln\frac{\mathcal{M}_0}{\mu^2}
\nn\\
&
\quad
+
\frac{\xi \kappa_0}{12v^2M^2_h}
\biggl[
\frac{12\mathcal{M}_0\mathcal{M}^2_1\mathcal{M}_2-2\mathcal{M}^4_1}{\mathcal{M}^2_0}
+3(3\mathcal{M}^2_2-3\mathcal{M}^2_1+4\mathcal{M}_1\mathcal{M}_3)
\nn\\
&
\qquad
\qquad
\qquad
\qquad
\qquad
\qquad
\qquad
\qquad
+\mathcal{M}_0(3\mathcal{M}_1-3\mathcal{M}_2+\mathcal{M}_4)
\biggr]
\,.
\end{align}
We note that,
in the SM, $\kappa_V=\kappa_{VV}=\kappa_3=\kappa_4=1$, $\kappa^{ij}_{u}=\kappa^{ij}_{d}=\kappa^{ij}_{e}=\delta^{ij}$ and 
the interactions with more than five fields vanish at the leading order.

We remark that 
$\kappa_3$ and $\kappa_4$ explicitly depend on the scale $\mu$. 
In the same way one can find that the coupling parameters for $\hat{h}^n~(n\geq 5)$ also depend on $\mu$. 
However, as we mentioned before, the physical coupling parameters should not depend on the scale $\mu$ after we determine the vacuum and physical Higgs mass. 
Assuming that $\mathcal{F}(h)$, $\mathcal{K}(h)$, $\mathcal{Y}(h)$,
and $\hat{\mathcal{Y}}(h)$ do not depend on the scale $\mu$,
we find that, to vanish the $\mu$ dependence in the renormalized Higgs potential, $\mathcal{M}_n~(n\geq 3)$ should satisfy
\begin{align}
\mathcal{M}_n
\,=\,
(\mathcal{M}_1-\mathcal{M}_2)
\,
\mathcal{G}_n(\mathcal{M}_0,\mathcal{M}_1,\mathcal{M}_2)\,,
\qquad (n\geq 3)
\,,
\label{eq:vanishingmu}
\end{align} 
where $\mathcal{G}_n$ are multivariable polynomials in $\mathcal{M}_{0}$, $\mathcal{M}_{1}$ and $\mathcal{M}_{2}$.
The multivariable polynomials for the first few orders are given as
\begin{align}
\mathcal{G}_3
&\,=\,
-3\,\frac{\mathcal{M}_0-\mathcal{M}_1}{\mathcal{M}_0}
\,,\\
\mathcal{G}_4
&\,=\,
-3\,\frac{\mathcal{M}^2_0-4\mathcal{M}^2_1-5\mathcal{M}_0\mathcal{M}_1-\mathcal{M}_0\mathcal{M}_2}{\mathcal{M}^2_0}
\,,\\
\mathcal{G}_5
&\,=\,
15\,
\frac{
4\mathcal{M}^3_0
+
\mathcal{M}^2_0\mathcal{M}_1
+
2\mathcal{M}^2_0\mathcal{M}_2
-
5\mathcal{M}_0\mathcal{M}^2_1
-
3
\mathcal{M}_0\mathcal{M}_1\mathcal{M}_2
}{\mathcal{M}^3_0}\,,\\
&
\qquad
\qquad
\qquad
\qquad
\qquad
\qquad
\vdots
\nn
\end{align}

Hereafter, we focus on a simple case where 
\begin{align}
\mathcal{M}_1=\mathcal{M}_2,\qquad
\mathcal{M}_n=0\quad(n\geq 3)
\,.
\label{eq:simple-Meq}
\end{align}
Clearly, Eq.~(\ref{eq:vanishingmu}) is satisfied in this case.
The polynomial $\mathcal{M}^2(h)$ is given in this case as
\begin{align}
\mathcal{M}^2(h)
&\,=\,
\mathcal{M}_0
+
\mathcal{M}_1\frac{h}{v}
+
\mathcal{M}_1\frac{h^2}{2v^2}
\nn\\
&\,=\,
\Lambda^2
+
\kappa_{\rm{p}}\l(
|\Phi|^2-\frac{v^2}{2}
\r)
\,,
\label{eq:simple-M}
\end{align}
where $\Phi$ denotes the SM Higgs doublet field,
\begin{align}
\Phi
\,=\,
\frac{1}{\sqrt{2}}
\,
(v+h)\,U
\l(
\begin{array}{ccc}
0\\
1
\end{array}
\r)\,,
\label{eq:Phi}
\end{align}
and $\kappa_{\rm{p}}$ is defined as
\begin{align}
\kappa_{\rm{p}}
=
\frac{\mathcal{M}_1}{v^2}
\,.
\end{align}
The parameter $\kappa_{\rm{p}}$ can be regarded as the portal coupling between the integrated new particle and the SM Higgs field. 
It is convenient to introduce
\begin{align}
r
\,=\,\frac{\frac{\kappa_{\rm{p}}}{2}v^2}{\Lambda^2}
\,=\,
\frac{\mathcal{M}_1}{2\mathcal{M}_0}\,,
\end{align}
which parameterizes ``non-decouplingness'' of the integrated new particles.
A non-decoupling case corresponds to the case with $|r|\simeq \mathcal{O}(1)$.
$\Delta\kappa_3$ and $\Delta\kappa_4$ are given in this case as
\begin{align}
\Delta\kappa_{3}
&\,=\,
-\frac{\xi}{2} (k_0+k_1)
+
\frac{4}{3}\,\xi\,
\frac{\mathcal{M}^2_0}{v^2 M^2_h}
\,\kappa_0\,r^3
\,,\\
\Delta\kappa_4
&\,=\,
-\xi\l(k_0+\frac{2}{3}k_2\r)
+
\frac{16}{3}\,\xi\,
\frac{\mathcal{M}^2_0}{v^2 M^2_h}
\,\kappa_0\,
\frac{r^3(3-r)}{2}
\,.
\end{align}
Using Eqs.~(\ref{eq:kappaV-naHEFT}) and (\ref{eq:kappaVV-naHEFT}),
we find
\begin{align}
\Delta\kappa_{3}
&\,=\,
\Delta\kappa_{VV}
-\Delta\kappa_{V}
+\frac{\xi}{2} (f_1-f_2)
+
\frac{4}{3}\,\xi\,
\frac{\mathcal{M}^2_0}{v^2 M^2_h}
\,\kappa_0\,r^3
\,,
\label{eq:kappa3-naHEFT}
\\
\Delta\kappa_4
&\,=\,
2(3-r)\Delta\kappa_3
+
2(4-r)\Delta\kappa_V
-
2(3-r)\Delta\kappa_{VV}
\nn\\
&
\qquad \qquad \qquad
-\xi\biggl[
(3-r)(f_1-f_2)
+f_1+\frac{2}{3}k_2
\biggr]
\,.
\label{eq:kappa4-naHEFT}
\end{align}

\subsection{Matching with extended scalar model}
\label{sec:model-scalar}
In order to confirm that the naHEFT describes the non-decoupling effects correctly,
let us compare the naHEFT and a concrete BSM scenario.
We consider an extension of the SM with $N$ real gauge singlet scalar bosons,
\begin{align}
\vec{S}\,=\,
(S_1,S_2,\cdots S_{N})\,,
\end{align}
where the scalar potential is given as
\begin{align}
V(\Phi,\vec{S})
\,=\,
m^2|\Phi|^2
+\lambda\, |\Phi|^4
+\frac{M^2}{2}(\vec{S}\cdot \vec{S})
+\frac{\kappa_{\rm{p}}}{2}|\Phi|^2(\vec{S}\cdot \vec{S})
+\frac{\lambda_S}{4}(\vec{S}\cdot \vec{S})^2
\,,
\end{align}
with $\Phi$ being the SM Higgs doublet field (Eq.~(\ref{eq:Phi})).
We take $\lambda>0$ and  $\lambda_S>0$ so that the scalar potential is bounded from below. 
The vacuum configuration of this model is obtained by solving the following couple of equation,
\begin{align}
\frac{\partial V}{\partial \phi}
&\,=\,
\phi\biggl[
m^2 +\lambda \phi^2 +\frac{\kappa_{\rm{p}}}{2}S^2_i
\biggr]
\,=\,
0\,,\\
\frac{\partial V}{\partial S_i}
&\,=\,
S_i\biggl[
M^2 +\lambda_S S^2_i +\frac{\kappa_{\rm{p}}}{2}\phi^2
\biggr]
\,=\,
0\,,
\label{eq:vacS}
\end{align}
where $\Phi=\frac{\phi}{2}\,U\,(0\,1)^T$.
We here assume $M^2\geq 0$ and $\kappa_{\rm{p}}\geq0$ so that $S_i$ do not obtain the VEVs,
\begin{align}
\braket{S_i}\,=\,0
\,.
\end{align}
The new scalar bosons $S_i$ do not mix with the SM Higgs field in this case.

The new scalar bosons $S_i$ are degenerate with the universal mass, $\Lambda$. 
The relation between $\Lambda$ and the parameters in the scalar potential is given as
\begin{align}
\Lambda^2=M^2+\frac{\kappa_{\rm{p}}}{2}v^2
\,.
\end{align}
We assume $\Lambda>v$ to work the EFT description in the later discussion.
For future convenience, we introduce a parameter $r$ as
\begin{align}
r
\,=\,
\frac{\frac{\kappa_{\rm{p}}}{2}v^2}{\Lambda^2}
\,=\,
1-\frac{M^2}{\Lambda^2}
\,,
\end{align}
which takes the value between 0 and 1 because of $\kappa_{\rm{p}}\geq 0$ and $\Lambda^2\geq M^2$. 
When $r\simeq 1$, the extra scalars acquire their mass mainly from the Higgs VEV. 

Integrating out these new scalars, we obtain the effective field theory with the form of (\ref{eq:naHEFT}). In this case the polynomials are determined as
 \cite{Henning:2014wua, Cohen:2020xca,Banta:2021dek}{\footnote{
 If we allow the mixing between $S_i$ and the SM Higgs field, the polynomials $\mathcal{F}$, $\mathcal{Y}$, and  $\hat{\mathcal{Y}}$ are nonzero.
 }}
\begin{align}
\mathcal{M}^2(h) 
&\,=\, 
\Lambda^2+\kappa_{\rm{p}}\l(
|\Phi|^2-\frac{v^2}{2}\r) 
\,,
\label{eq:HSM-M}
\\
\mathcal{K}(h) 
&\,=\,
\kappa_2\,
\frac{\Lambda^2}{3v^2}
\,r\,
\biggl[
1-\frac{M^2}{\mathcal{M}^2(h)}
\biggr]
\,,
\label{eq:HSM-K}
\\
\mathcal{F}(h) 
&\,=\,
\mathcal{Y}^{ij}(h) \,=\,
\hat{\mathcal{Y}}^{ij}(h) \,=\,0\,,
\label{eq:HSM-zeros}
\end{align}
and $\kappa_0 = \kappa_2=N$.
We note that this setup satisfies Eq.~(\ref{eq:simple-Meq}).
The relation between $h$ and the canonical Higgs field ($\hat{h}$) (up to $\mathcal{O}(\xi^2)$) is calculated as
\begin{align}
h
=
\hat{h}
-\kappa_2\,\xi\, \frac{\Lambda^2}{6v^2}r\,
\biggl[\hat{h}
-\,(1-r)\,v \,\mathcal{H}(\hat{h}/v)
\biggr]
\,,
\end{align}
where
\begin{align}
\mathcal{H}(x)
&\,=\,
\frac{\tan^{-1}\l(\sqrt{\frac{r}{1-r}}\l(1+x\r)\r)-\sin^{-1}\sqrt{r}}{\sqrt{r(1-r)}}
\,=\,
x-r x^2+\mathcal{O}(x^3)
\,.
\end{align}
It is straightforward to confirm that $h=\hat{h}$ when $\mathcal{K}(h)=0$ (namely, $\kappa_2=0$).

\begin{figure}[t]
	\centering
	\includegraphics[width=8cm,clip]{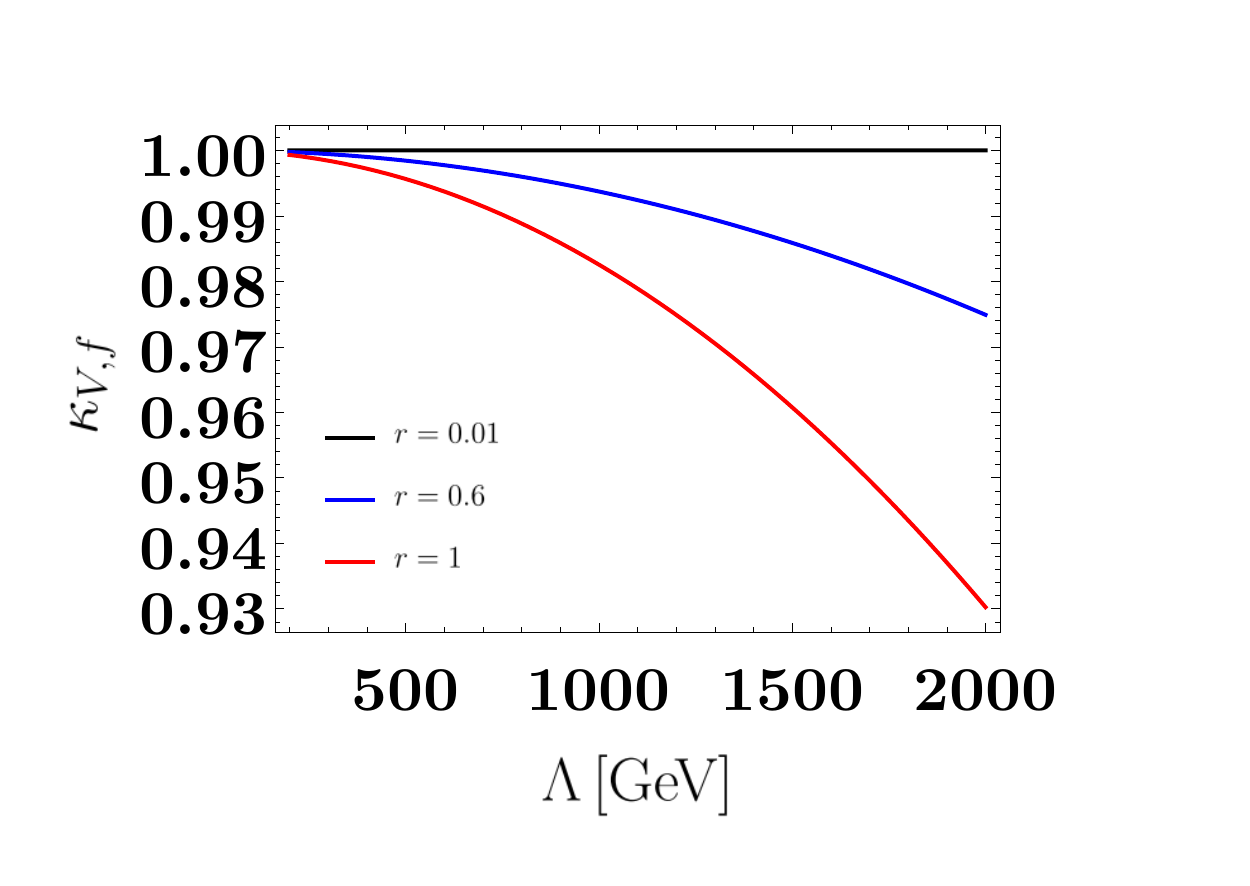}
	 \hspace{-1.5cm}
	\includegraphics[width=8cm,clip]{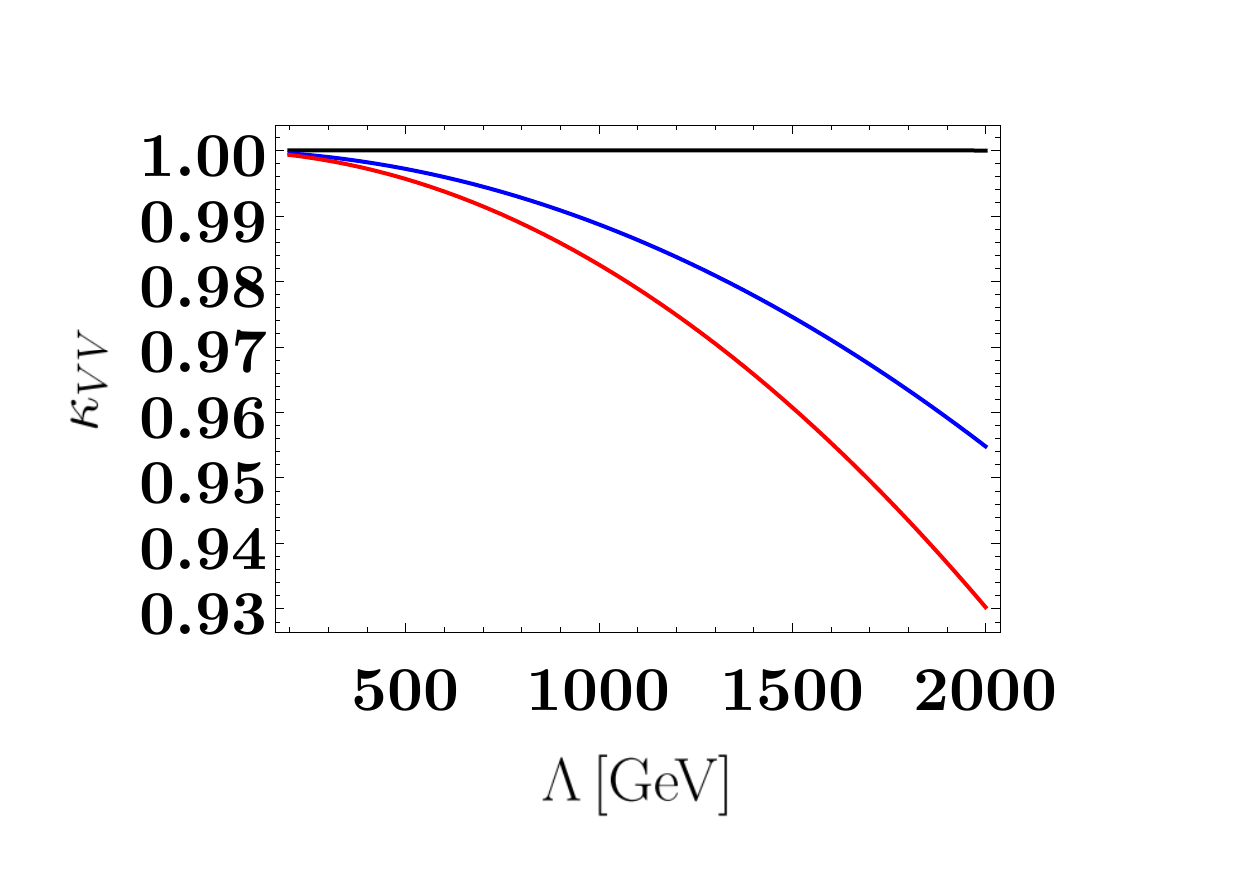}
	\\
	\includegraphics[width=8cm,clip]{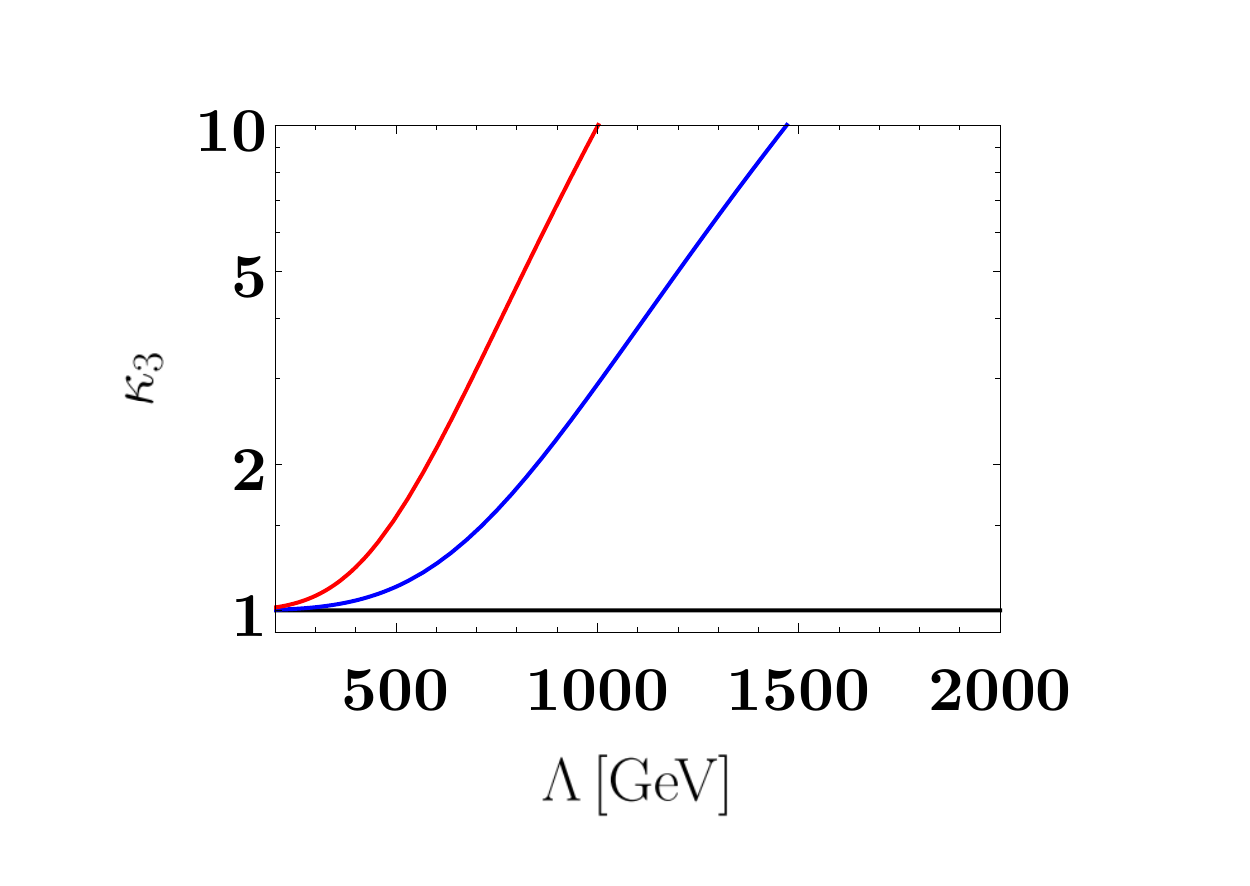}
	 \hspace{-1.5cm}
	\includegraphics[width=8cm,clip]{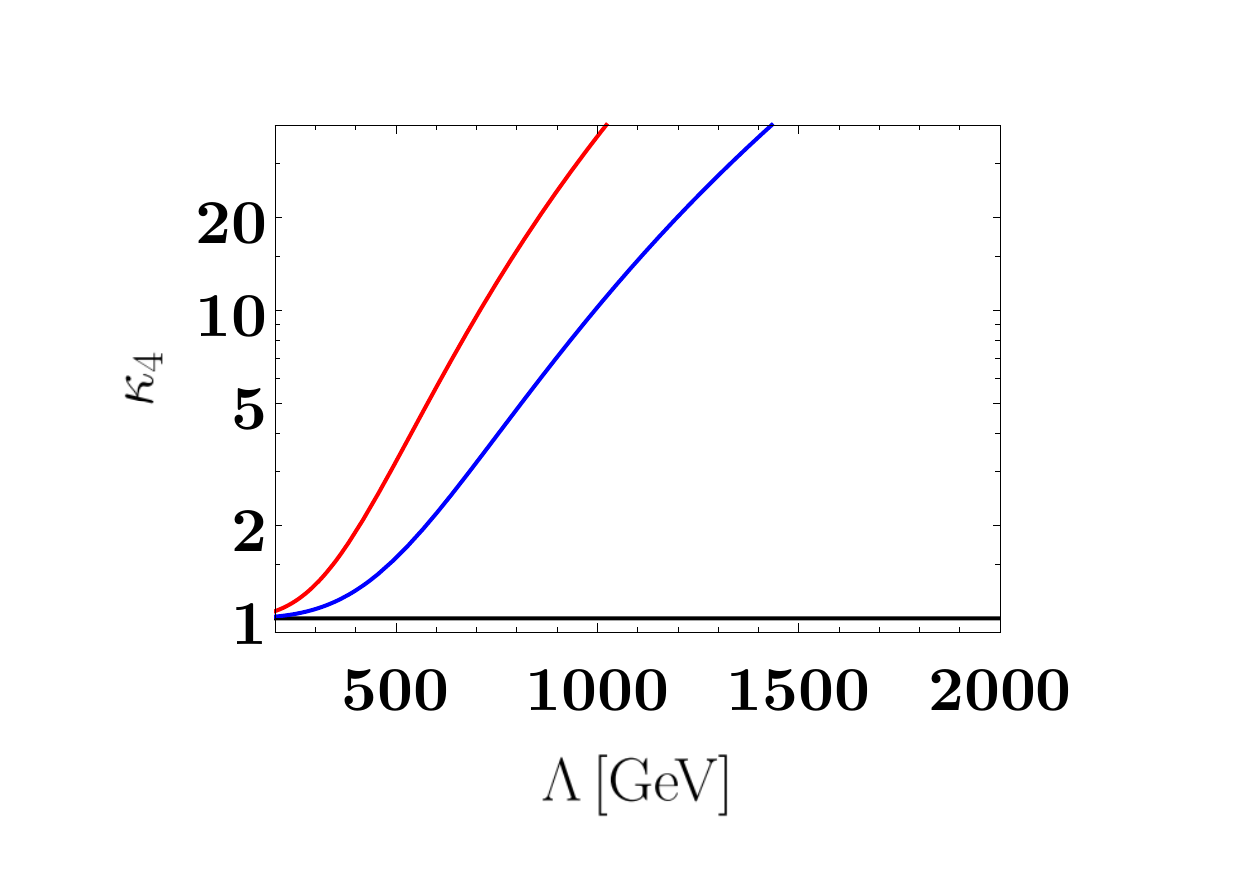}
		\caption{ 
		Higgs coupling factors for the model with $N=1$. 
				}
	\label{fig:higgcoup}
\end{figure}
The Higgs coupling factors in this case are obtained as
\begin{align}
\kappa_{V}
&\,=\,
1
-
\kappa_2\,
\frac{\xi}{6}
\,\frac{\Lambda^2}{v^2}{r^2}
\label{eq:kappaV-scalar}
\,,\\
\kappa^{ij}_{u}
&\,=\,
\kappa^{ij}_{d}
\,=\,
\kappa^{ij}_{e}
\,=\,
\kappa_V
\delta_{ij}
\,,\\
\kappa_{VV}
&\,=\,
1
-
\,
\kappa_{2}
\,
\frac{\xi}{6}
\frac{\Lambda^2}{v^2}
{r^2}
(3-2r)
\,,\\
\kappa_{3}
&\,=\,
1
+
\frac{4\xi}{3}
\frac{\Lambda^4}{v^2M^2_h} \,
\biggl[
\kappa_{0}
\,
r^3
-
\kappa_{2}
\,
\frac{M^2_h}{8\Lambda^2}
\,
r^2(3-2r)
\biggr]\,,
\label{eq:kappa3-scalar}\\
\kappa_{4}
&\,=\,
1+
\frac{16\xi}{3}
\frac{\Lambda^4}{v^2M^2_h} \,
\biggl[
\kappa_{0}
\,
r^3
\,
\frac{(3-r)}{2}
-
\kappa_{2}
\,
\frac{M^2_h}{16\Lambda^2}
\,
r^2
\frac{(25-38r+16r^2)}{3}
\biggr]
\,,
\label{eq:kappa4-scalar}
\end{align}
where we ignore the $\mathcal{O}(\xi^2)$ corrections.
In figure \ref{fig:higgcoup}, we estimate the Higgs coupling parameters for the $N=1$ case. Black, blue and red lines correspond to the cases with $r=0.01$, $0.6$, and $1$, respectively.  
We find that, if $r\simeq 0$, the coupling derivations are highly suppressed. On the other hand, if we take $r\simeq 1$, the coupling deviation can be sizable for a large $\Lambda$ due to the enhancement factor with the power of $\Lambda$. These enhancement corrections are regarded as non-decoupling effects from the integrated particles. 
Such a non-decoupling property has been pointed out in the concrete extended Higgs models in Refs.~\cite{
Kanemura:2002vm, 
Kanemura:2004mg
}. 
We thus find that the non-decoupling property can be effectively parameterized by the naHEFT.

\begin{figure}
	\centering
	\includegraphics[width=5.5cm,clip]{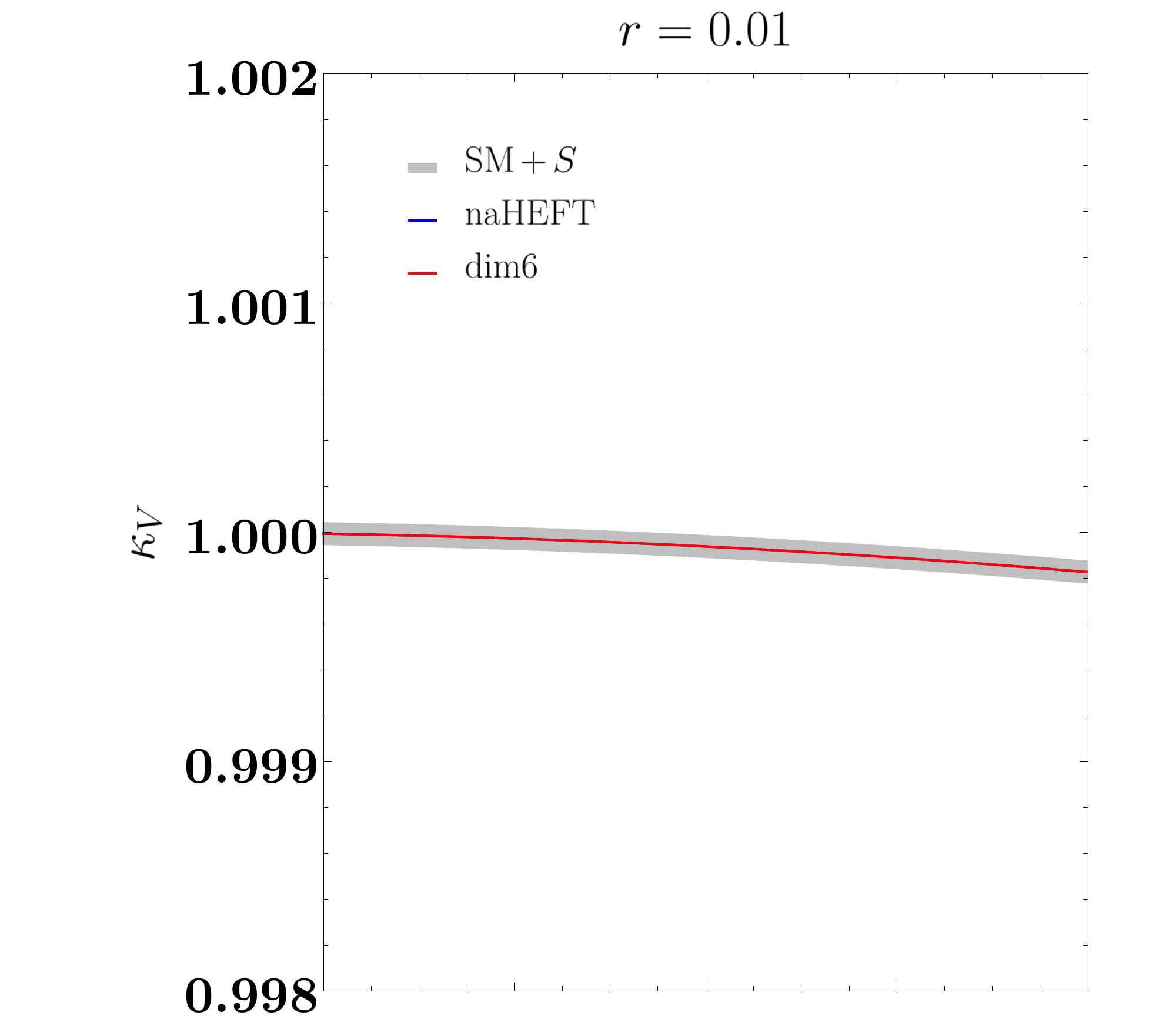}
	\hspace{-1.5cm}
	~~
	\includegraphics[width=5.5cm,clip]{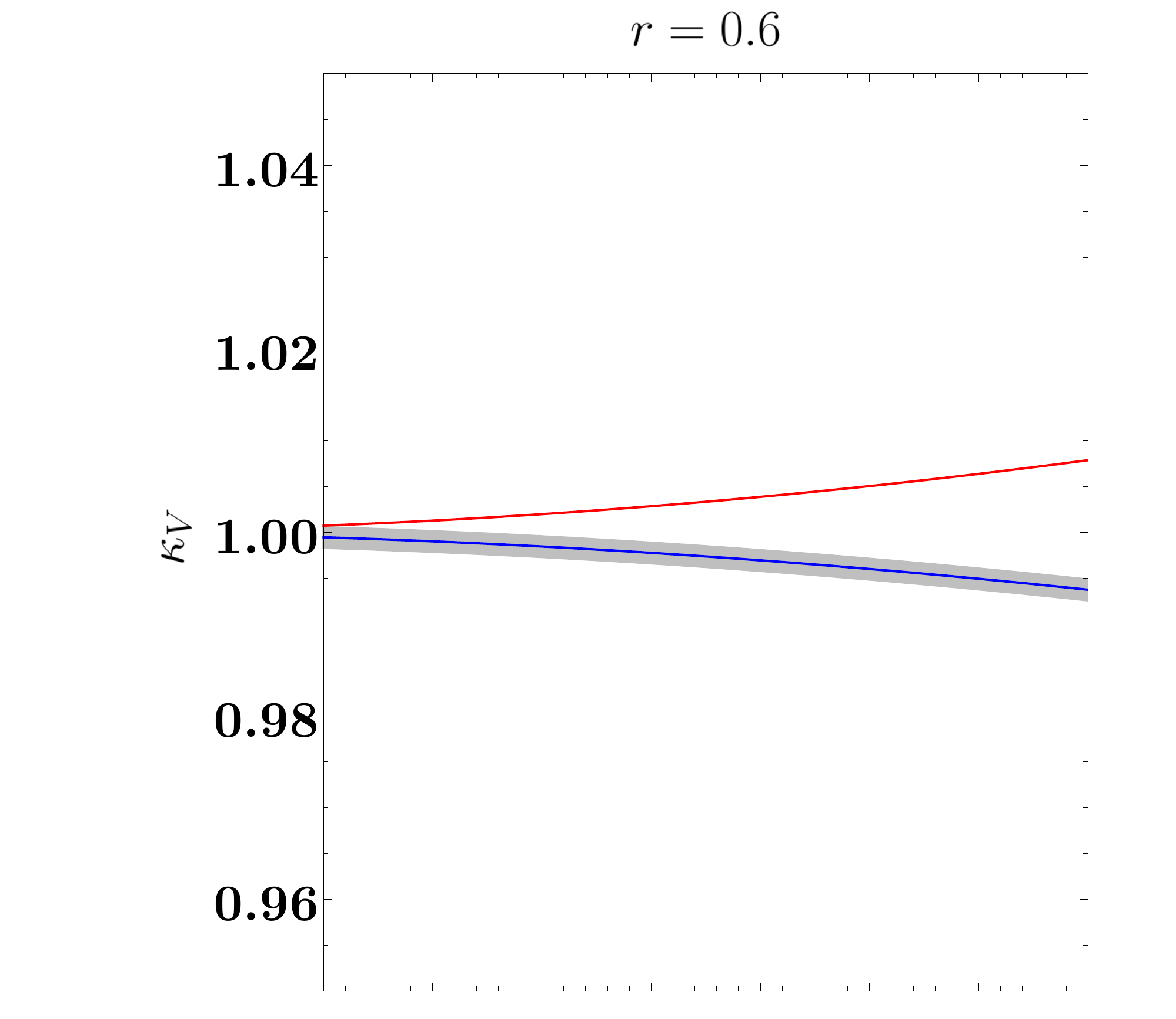}
	\hspace{-1.5cm}
	~~
	\includegraphics[width=5.5cm,clip]{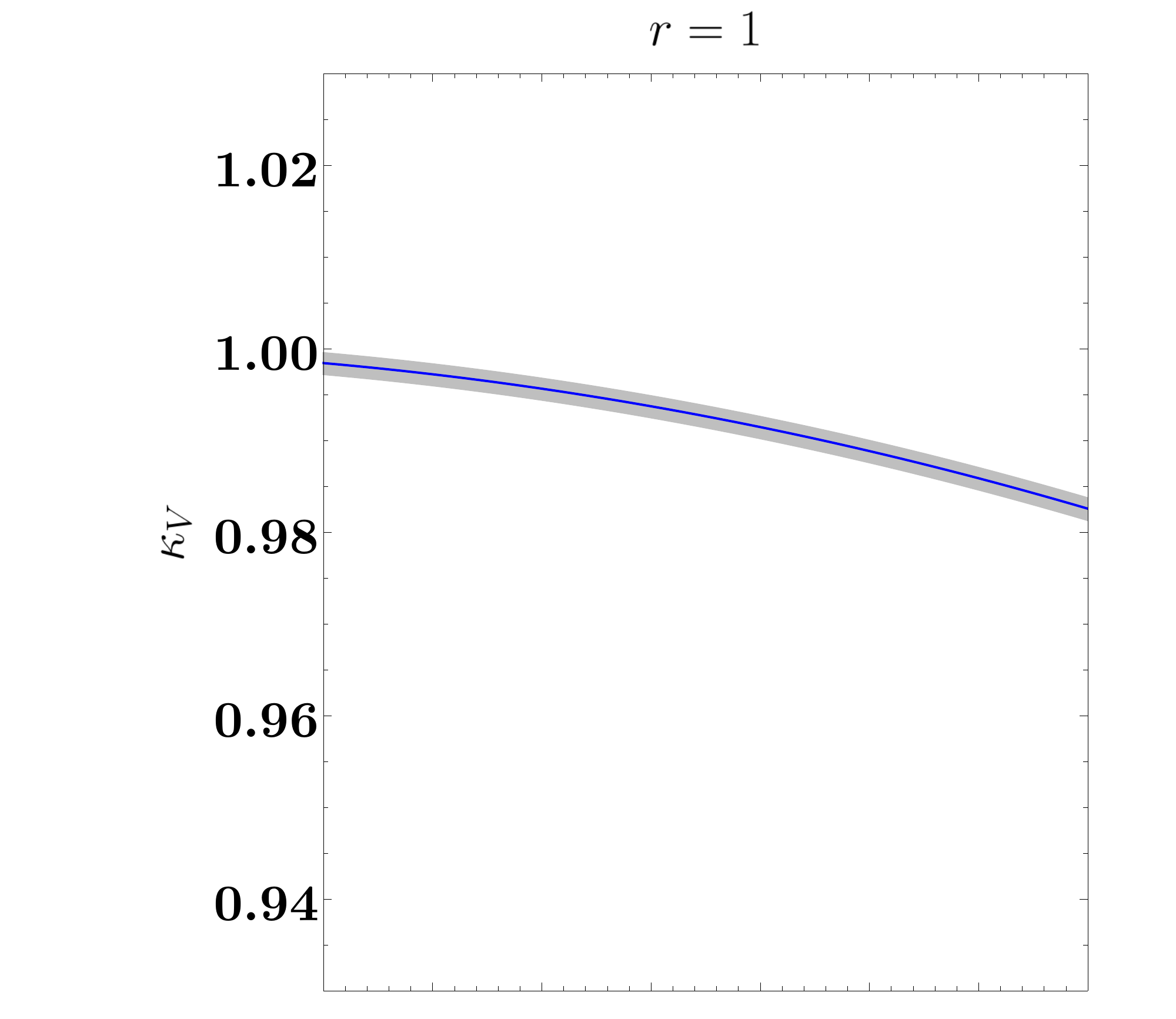}
	\\
	\includegraphics[width=5.5cm,clip]{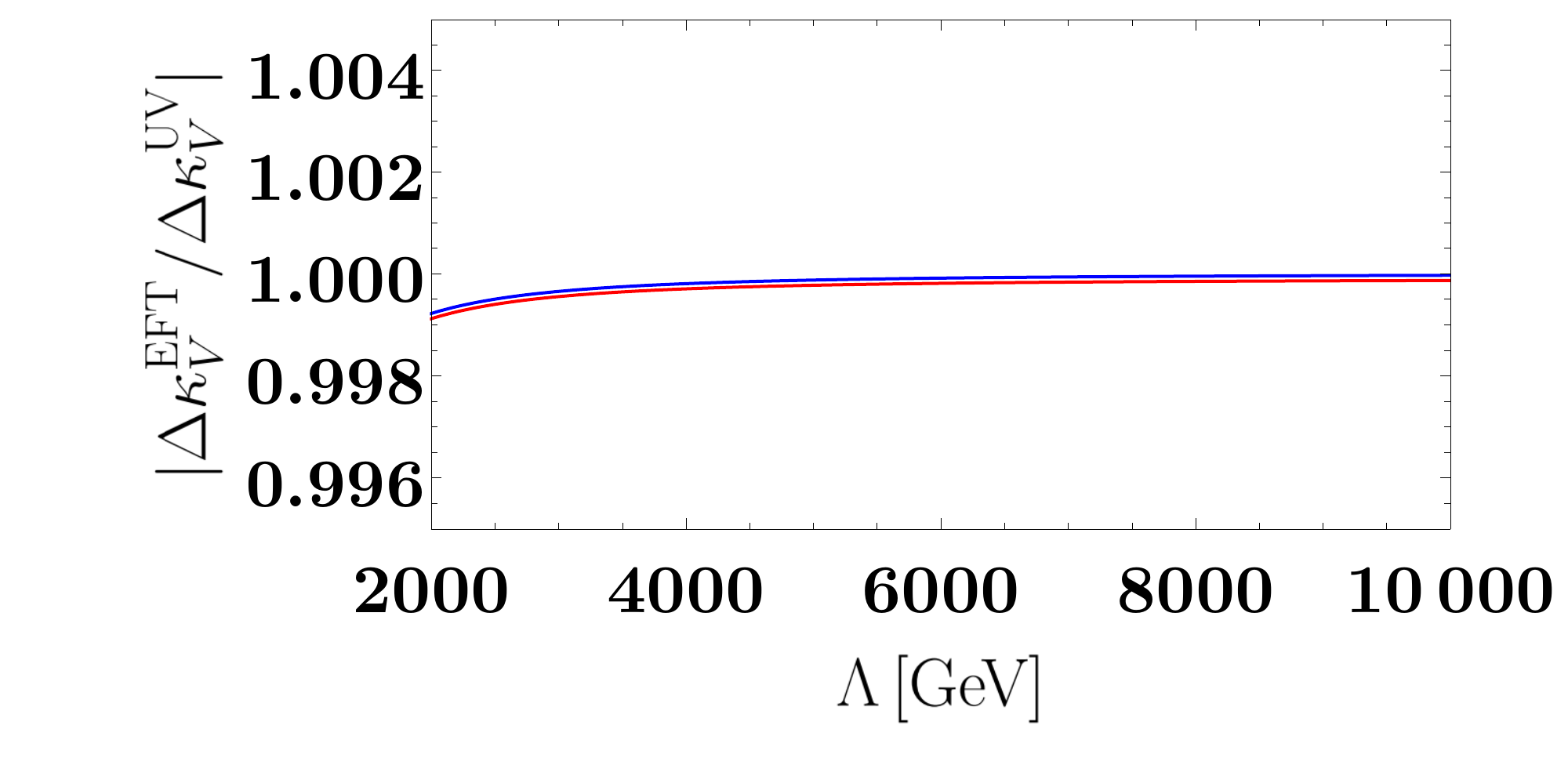}
	\hspace{-1.5cm}
	~~
	\includegraphics[width=5.5cm,clip]{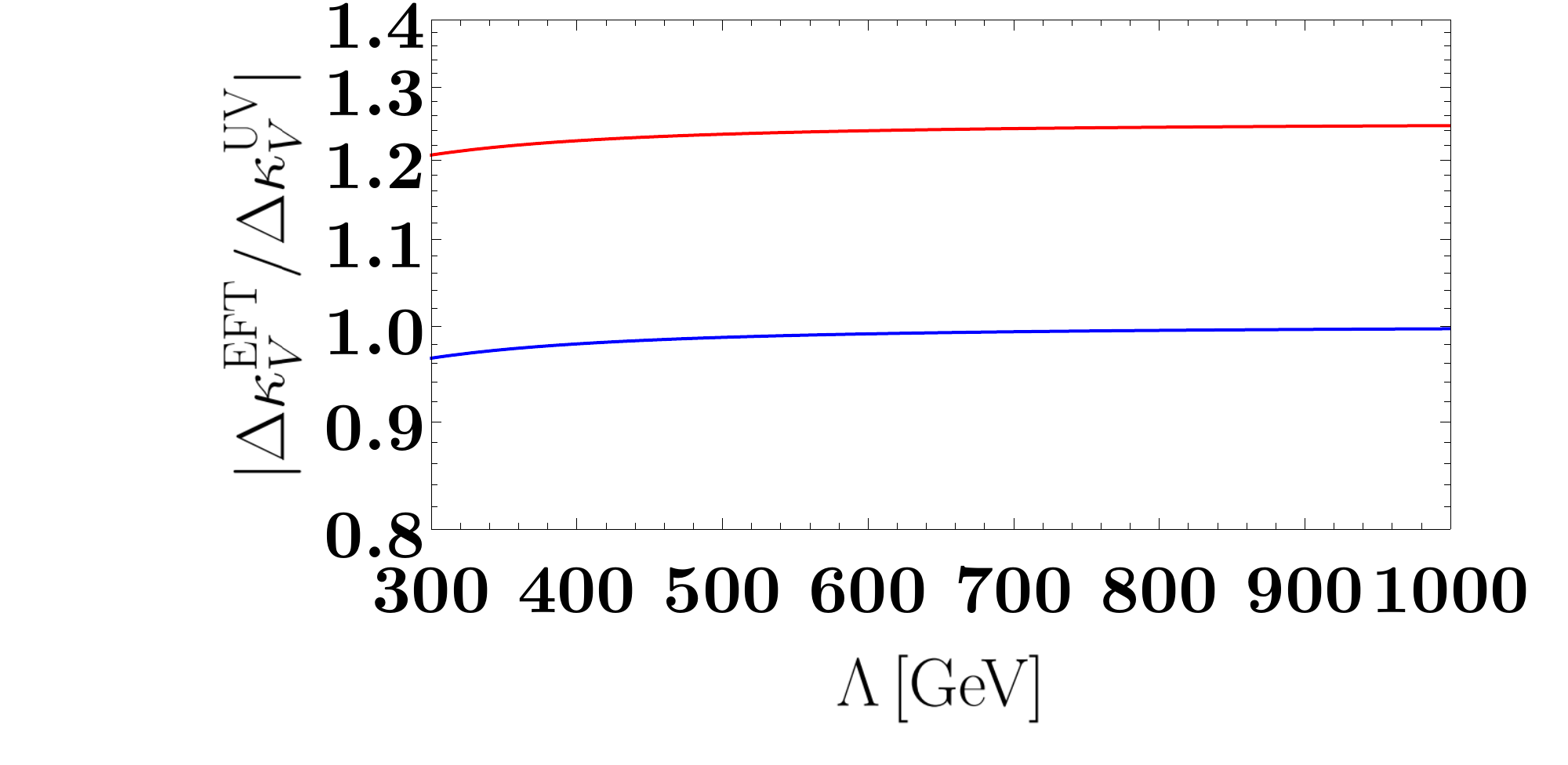}
	\hspace{-1.5cm}
	~~
	\includegraphics[width=5.5cm,clip]{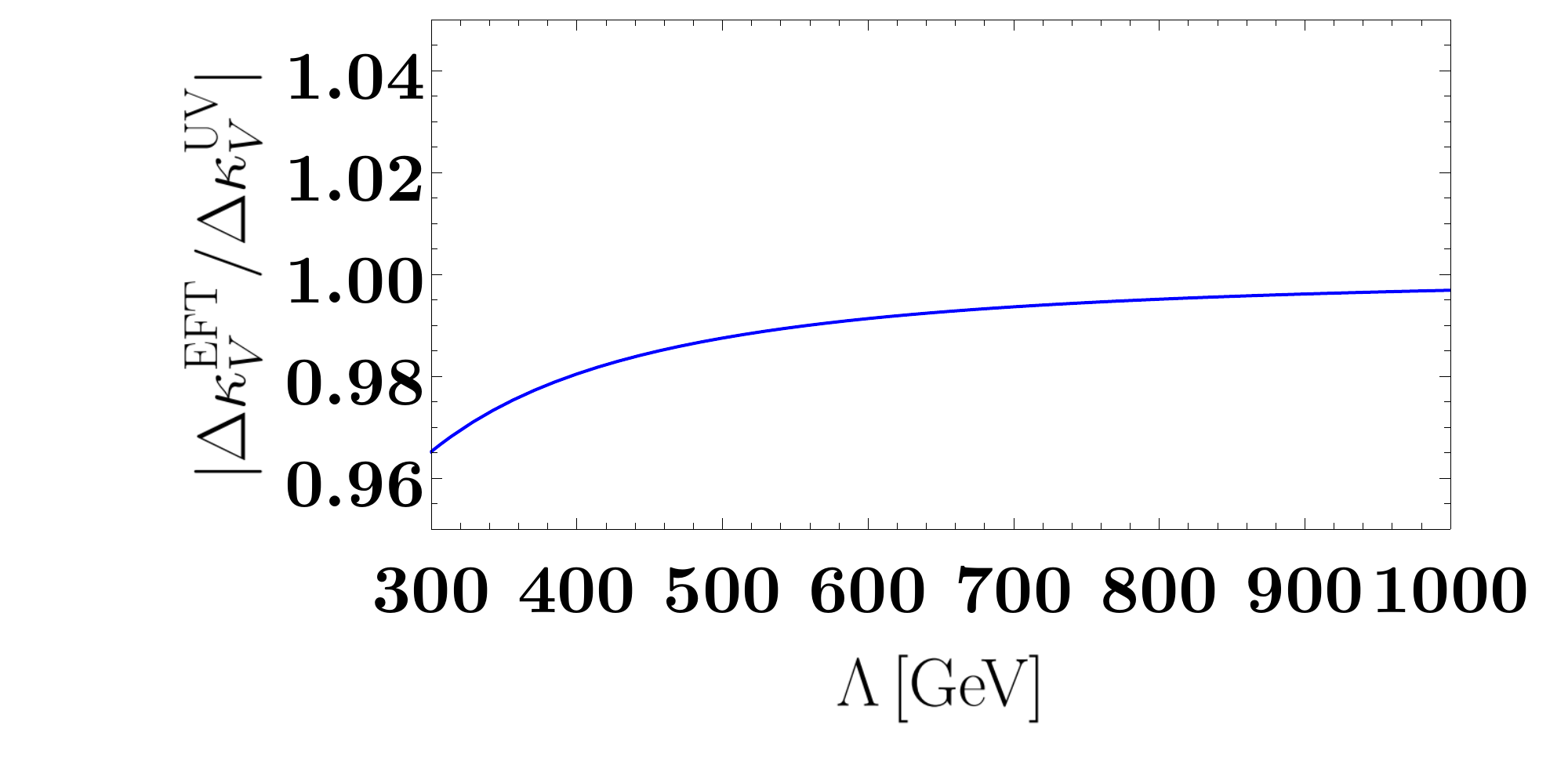}
	\\
	\vspace{1cm}
        \includegraphics[width=5.5cm,clip]{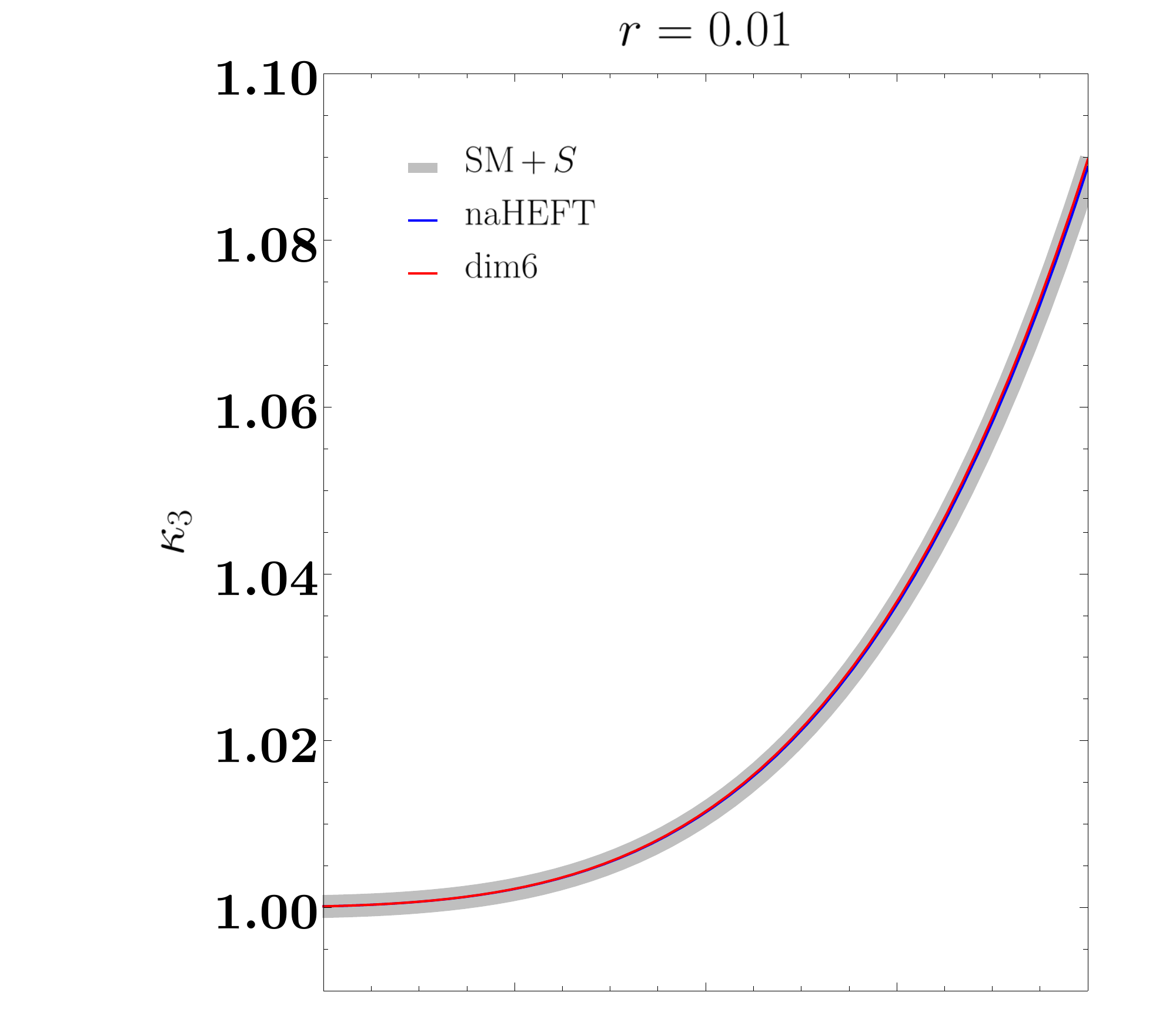}
	\hspace{-1.5cm}
	~~
	\includegraphics[width=5.5cm,clip]{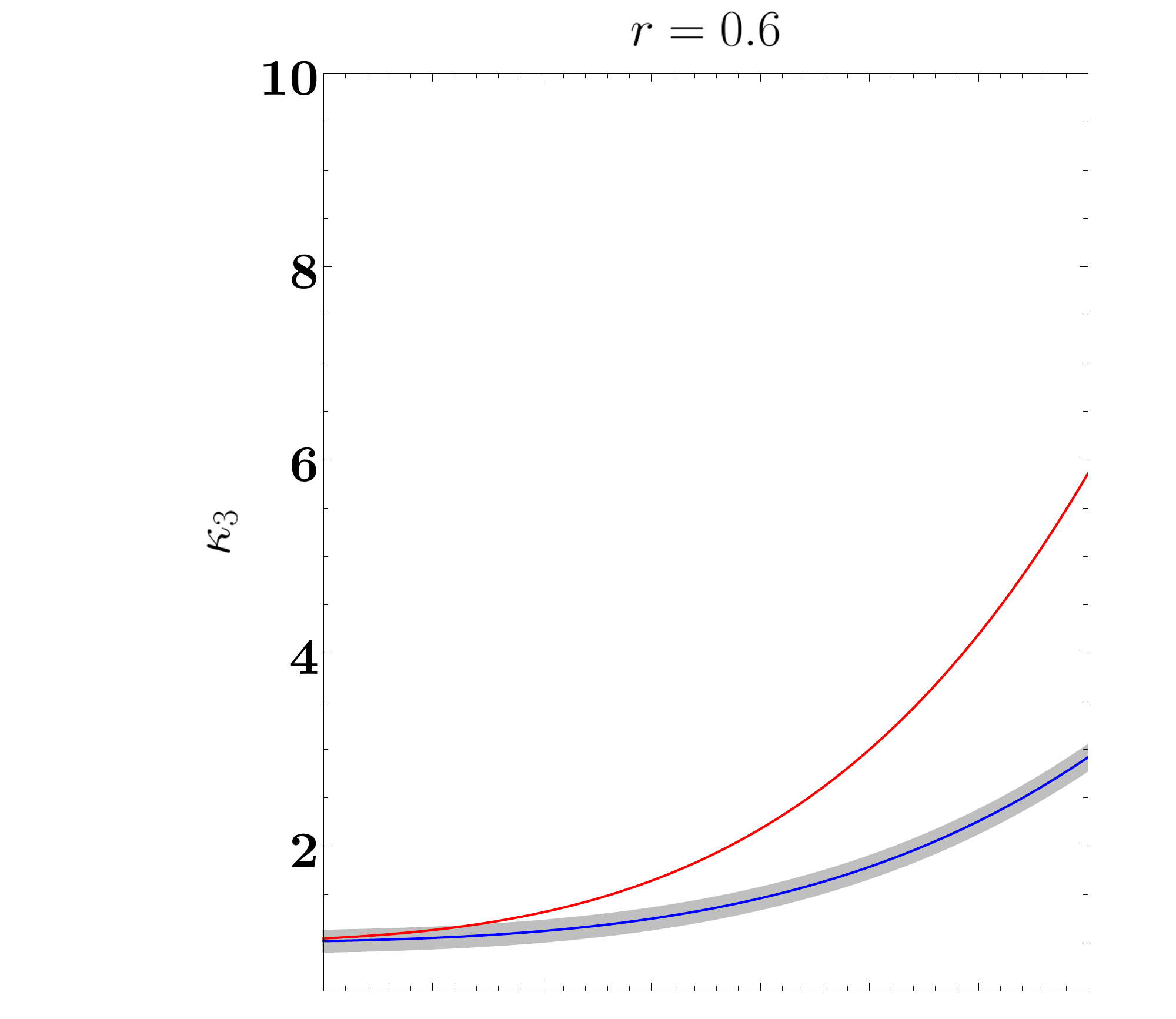}
	\hspace{-1.5cm}
	~~
	\includegraphics[width=5.5cm,clip]{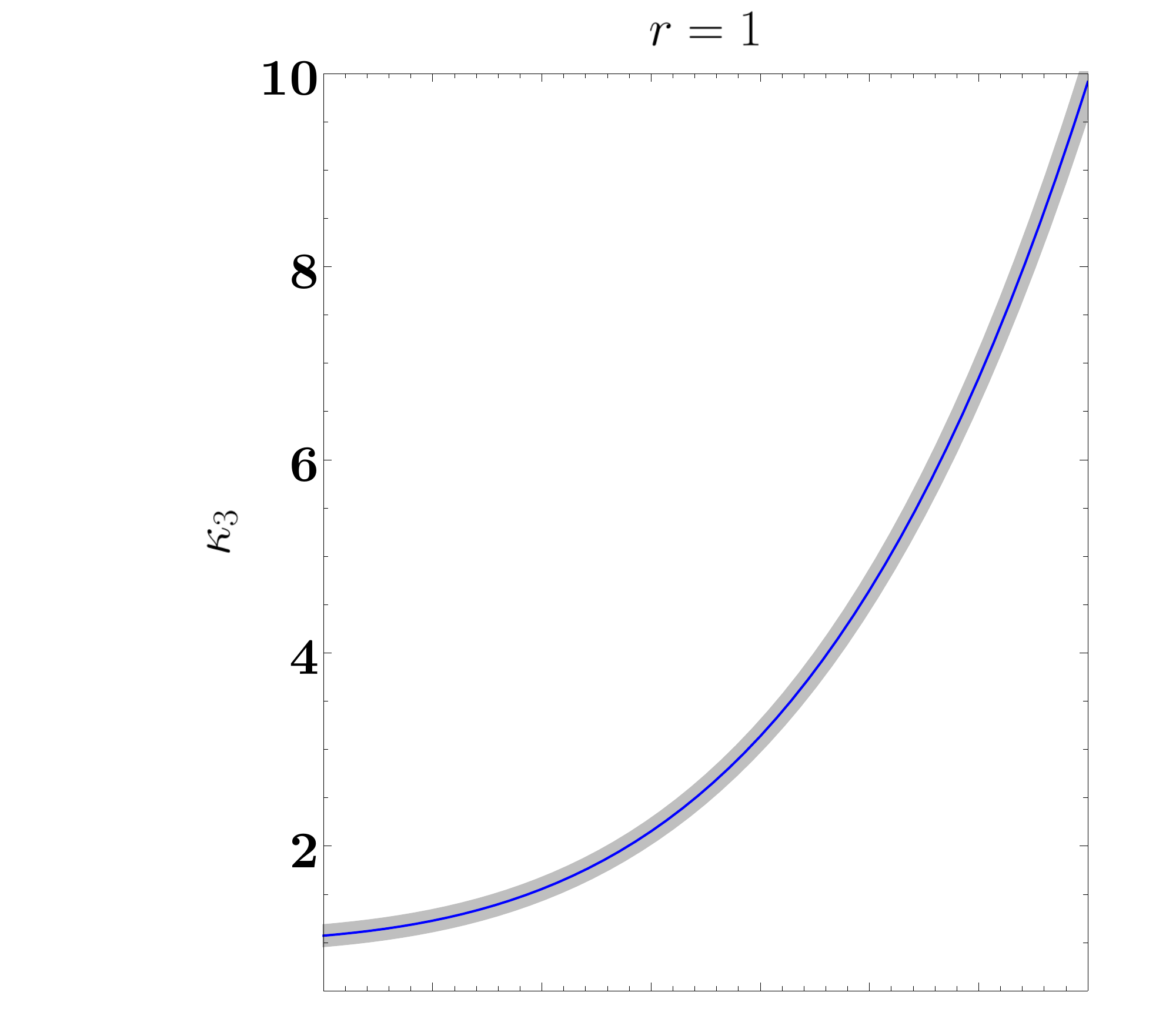}
	\\
	\includegraphics[width=5.5cm,clip]{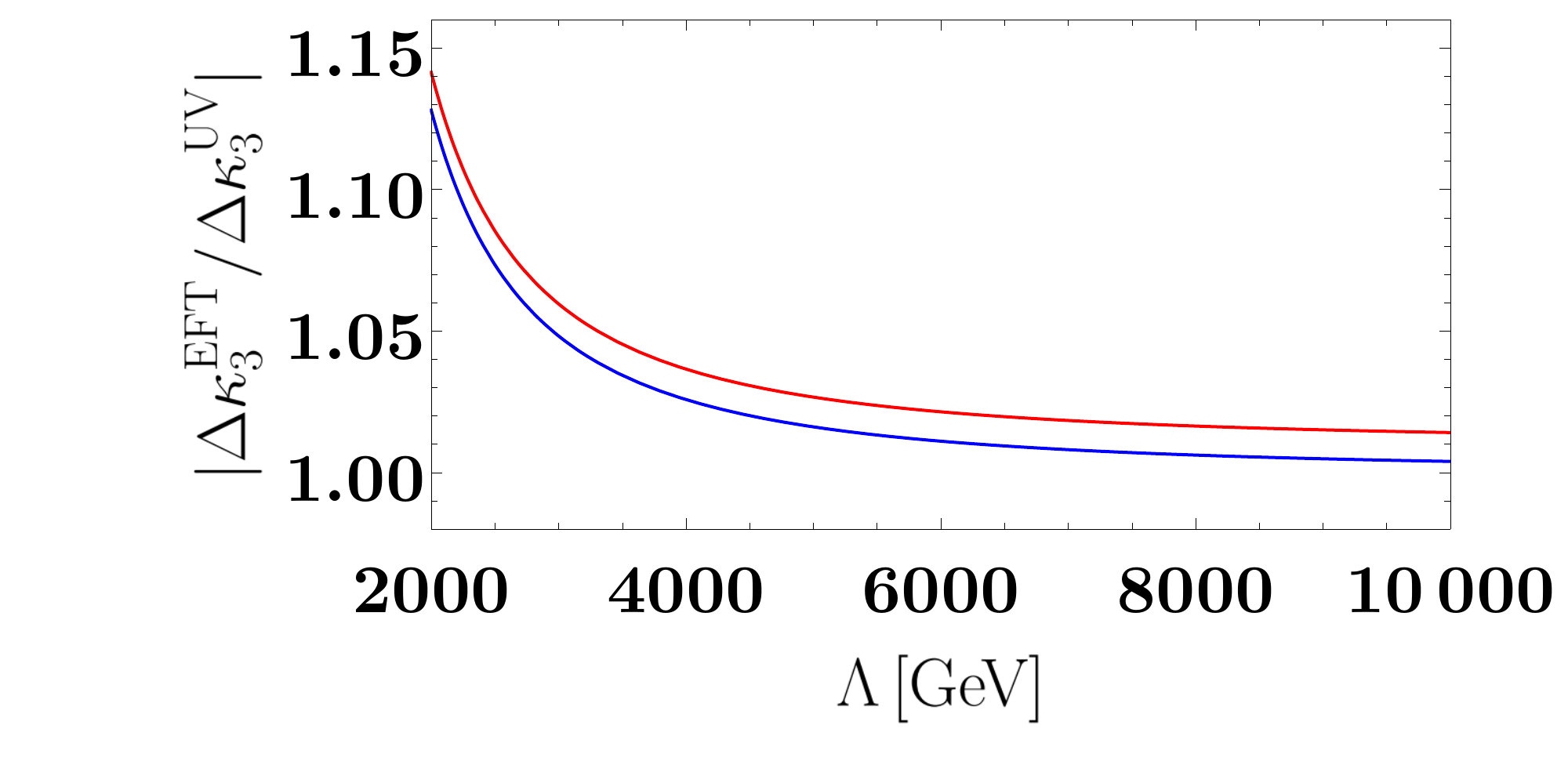}
	\hspace{-1.5cm}
	~~
	\includegraphics[width=5.5cm,clip]{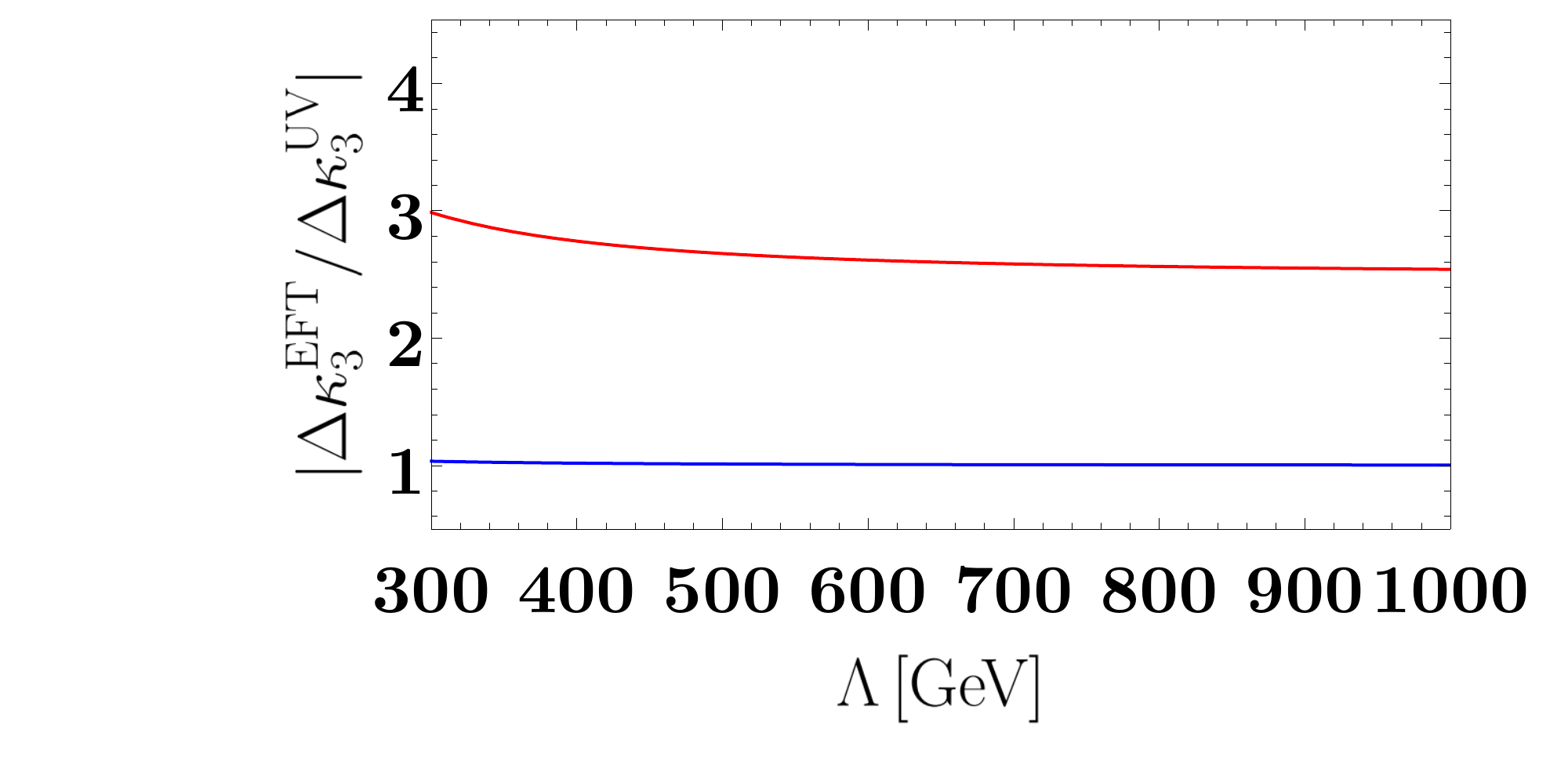}
	\hspace{-1.5cm}
	~~
	\includegraphics[width=5.5cm,clip]{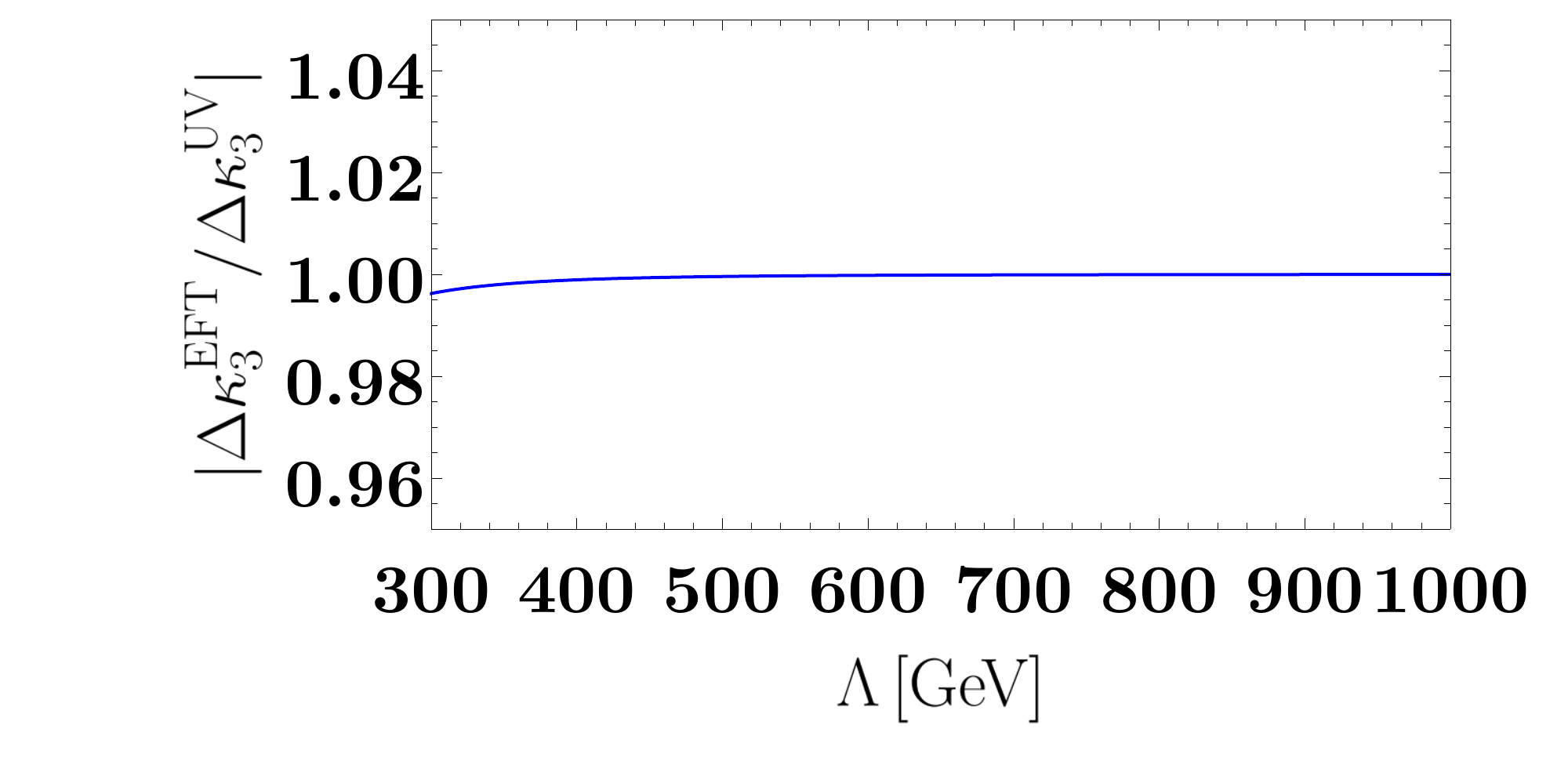}
		\caption{ 
		We compare the various EFT results with the full one-loop results in the renormalizable model reported in Refs.~\cite{Kanemura:2015fra, Kanemura:2016lkz}. $\Delta \kappa_{V,3} = \kappa_{V,3}-1$. In the full one-loop calculations, we only include the singlet-loop effects. 
				}
	\label{fig:UVmodel}
\end{figure}
We note that, if $r\neq 1~(M\neq 0)$, the naHEFT Lagrangian can be expressed in terms of the polynomial in $|\Phi|^2$, which is a familiar form called the ``Standard Model Effective Field Theory (SMEFT) \refsmeft''.
From Eqs.~(\ref{eq:HSM-M})-(\ref{eq:HSM-zeros}), we find 
\begin{align}
\mathcal{L}_{\rm{naHEFT}}
\,=\,
\mathcal{L}_{\rm{SM}}
&\,-\,
\frac{\xi}{4}
\kappa_0
\l(M^2+\kappa_{\rm{p}} |\Phi|^2\r)^2
\biggl[
\ln\frac{M^2}{\mu^2}
+\sum_{n=1}^{\infty}
\frac{(-1)^{n+1}}{n}
\l(\frac{\kappa_{\rm{p}}|\Phi|^2}{M^2}\r)^n
\biggr]
\nn\\
&
+
\frac{\xi}{24}
\kappa_2
\frac{\kappa^2_{\rm{p}}}{M^2}
\sum_{n=0}^{\infty}
\l(-\frac{\kappa_{\rm{p}}|\Phi|^2}{M^2}\r)^n
\,
\partial_\mu |\Phi|^2\,\partial^\mu |\Phi|^2
\,.
\end{align}
If we truncate the effective Lagrangian at mass dimension six, 
we obtain
\begin{align}
\mathcal{L}_{\rm{naHEFT}}|_{\rm{dim6}}
\,=\,
\mathcal{L}_{\rm{SM}}
&
\,-\,
\frac{\xi}{2}\,\kappa_{\rm{p}}\,
\biggl[
\kappa_{0}\l(\ln\frac{M^2}{\mu^2}+\frac{1}{2}\r)M^2|\Phi|^2
+\kappa_{0}\l(\ln\frac{M^2}{\mu^2}+\frac{3}{2}\r)\kappa|\Phi|^4
\nn\\
&
+\kappa_{0}\,
\frac{\kappa^2_{\rm{p}}}{6M^2}|\Phi|^6
-\kappa_{2}\,
\frac{\kappa_{\rm{p}}}{12M^2}\,
\partial_\mu |\Phi|^2\,\partial^\mu |\Phi|^2
\biggr]
\,.
\label{eq:Ldim6}
\end{align}
We note that the dimension six approximation works only for small $r$. For the case with $r\simeq 1$, effects of the higher-dimensional operators cannot be ignored. We will clarify this point below.

In order to confirm validity of our effective field theory description, 
we compare the Higgs coupling parameters estimated by i) naHEFT, ii) dimension six EFT, and iii) the complete one-loop calculations without integrating out the extra scalar bosons in the renormalizable model. 
In figure \ref{fig:UVmodel}, we plot $\kappa_V$ and $\kappa_3$ for the $N=1$ case. The gray line corresponds to the complete one-loop calculation of the extra scalar contribution in a real-singlet scalar extended model, which have been reported in Refs.~\cite{Kanemura:2015fra, Kanemura:2016lkz}. 
On the other hand, the blue and red lines are the results obtained from the naHEFT and dimension six EFT, respectively. 
We plot the ratio of the EFT results of $\Delta \kappa_{V,3} = \kappa_{V,3}-1$ to the full one-loop results in the bottom figures. We find that both the EFT results agree with the full one-loop results for $r\simeq 0$ case. 
However, if we take $r\simeq 1$, the dimension six approximation fails. 
This observation means that the dimension six EFT cannot be applied for the non-decoupling case. The invalidity of the dimension six EFT for the non-decoupling case has also pointed out in Refs.~\cite{Falkowski:2019tft,Cohen:2020xca,Cohen:2021ucp}. 
In contrast with the dimension six EFT, 
the naHEFT agrees with the complete one-loop calculations even for $r\simeq 1$. 
We thus confirm that the naHEFT can be applied for not only decoupling case ($r\simeq 0$) but also non-decoupling case ($r\simeq 1$). 

\section{Vacuum stability}
\label{sec:vacuum stability}
In the previous section, we assumed that the EW vacuum is the global minimum. 
However, this assumption is not always justified in the naHEFT framework because the Higgs potential is modified from the SM one.
In this section, we consider the vacuum structure in the naHEFT and 
discuss when the EW vacuum is ensured to be the global minimum.

The relevant part for the vacuum stability in naHEFT is 
\begin{align}
\mathcal{L}_{\rm{naHEFT}}
\,\supset\,
\frac{1}{2}\biggl(1+\xi \mathcal{K}(h)\biggr)(\partial_\mu h)(\partial^\mu h)
-V_{\rm{naHEFT}}(h)
\,,
\end{align}
where 
\begin{align}
V_{\rm{naHEFT}}
&\,=\,
\frac{m^2}{2}h^2
+
\frac{\lambda}{4}h^4
+
\frac{\xi}{4}\,\kappa_0\,
[\mathcal{M}^2(h)]^2
\ln \frac{\mathcal{M}^2(h)}{\mu^2}
\,.
\end{align}
The first two terms in $V_{\rm{naHEFT}}$ come from the SM Lagrangian, $\mathcal{L}_{\rm{SM}}$. $m^2$ and $\lambda$ are both real parameters. 
Here we ignore the one-loop corrections from the SM particles for simplicity{\footnote{We have confirmed that the SM loop corrections are subdominant in our numerical analysis performed below.}}. 
We estimate the vacuum stability of the naHEFT as follows :
\begin{itemize}
\item Perform the field redefinition, (\ref{eq:htohhat})\,,
\item Introduce an order parameter, $\phi=v+\hat{h}$\,,
\item Impose that ``$\phi=v$ is a global minumum of $V_{\rm{naHEFT}}$ for $|\phi|<\Lambda$''.
\end{itemize}
We here focus on only the region where the field value is smaller than the cutoff scale of our EFT description, $\Lambda$.

\begin{figure}
	\centering
	\includegraphics[width=7cm,clip]{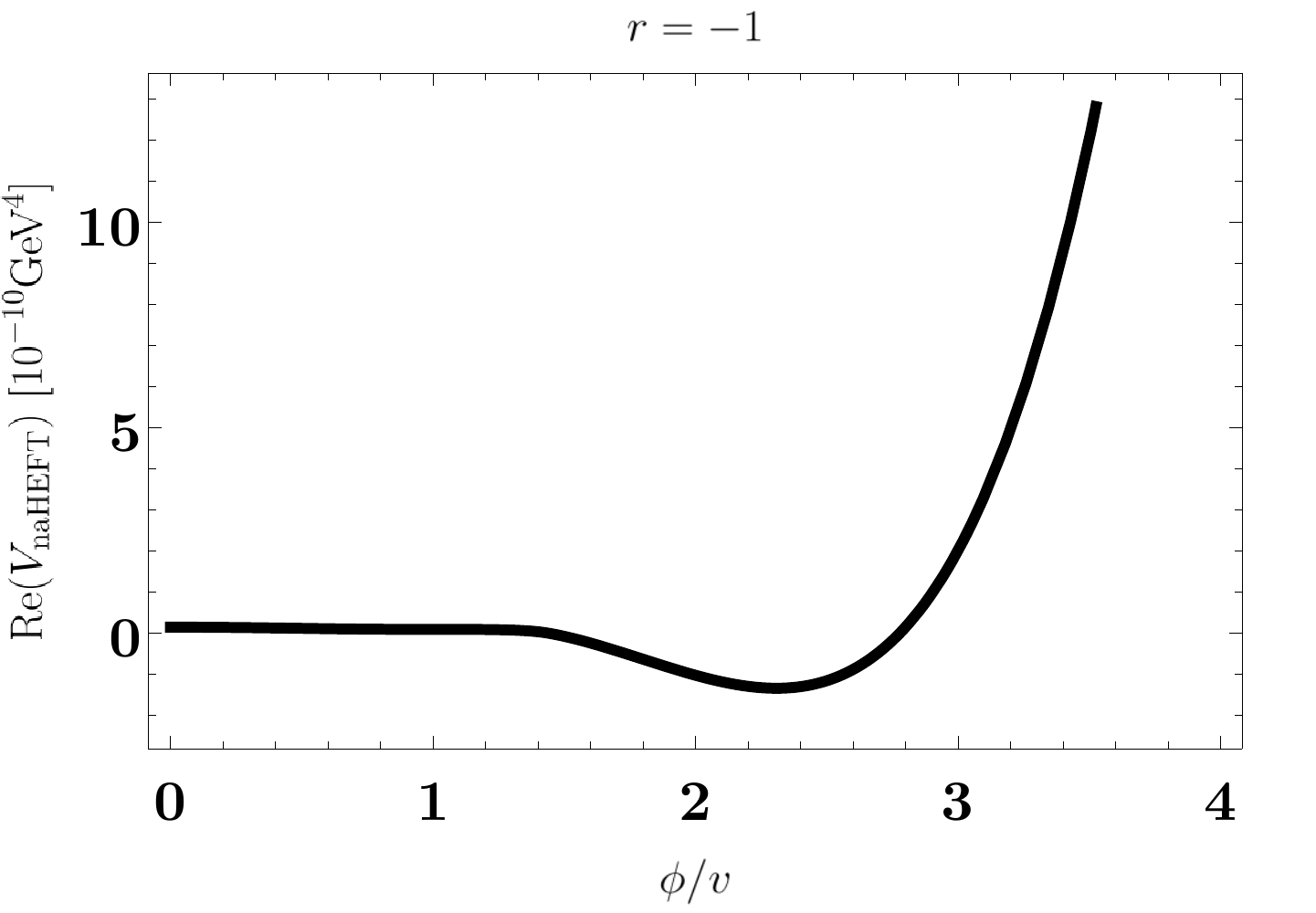}
	\includegraphics[width=7cm,clip]{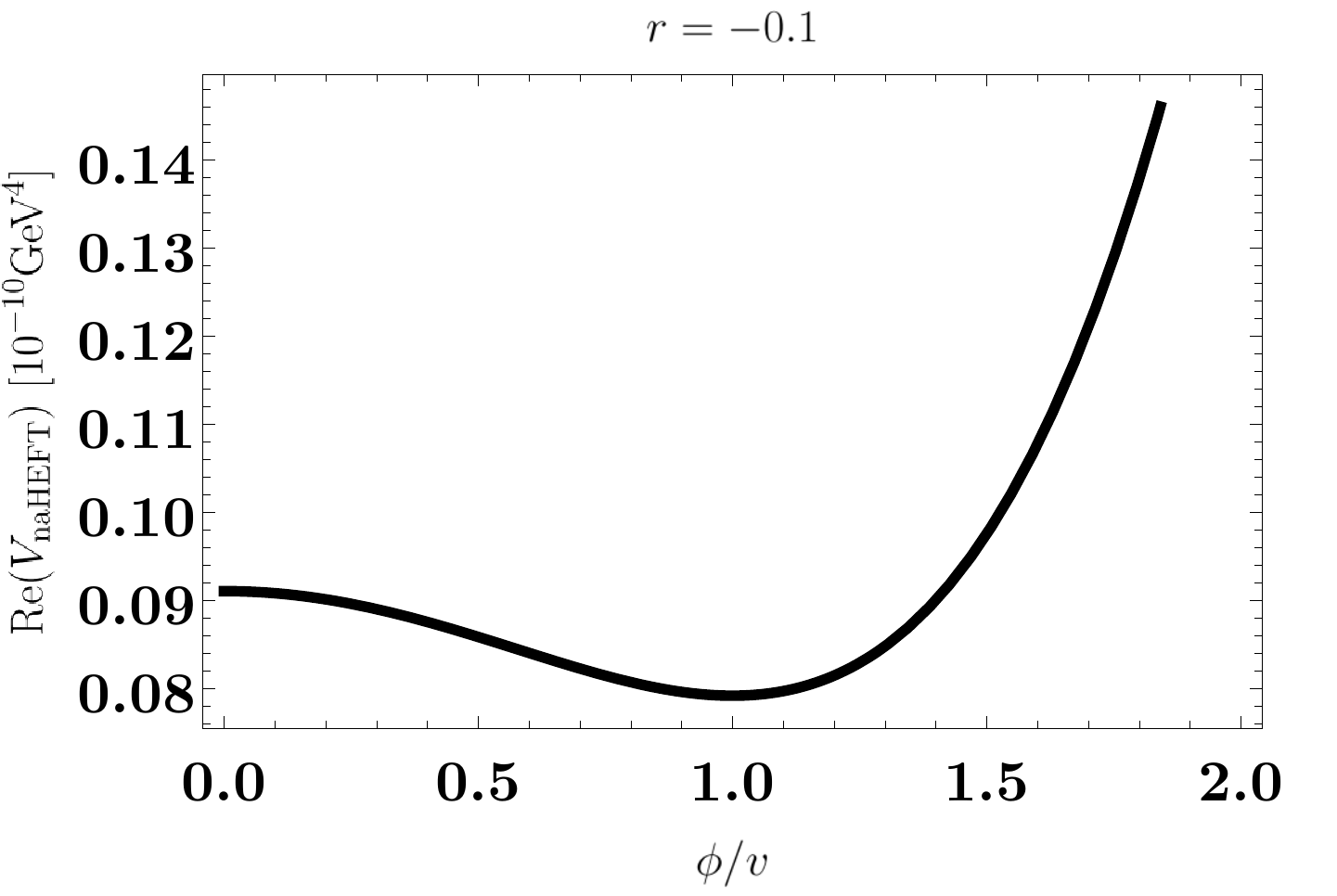}
	\\
	\includegraphics[width=7cm,clip]{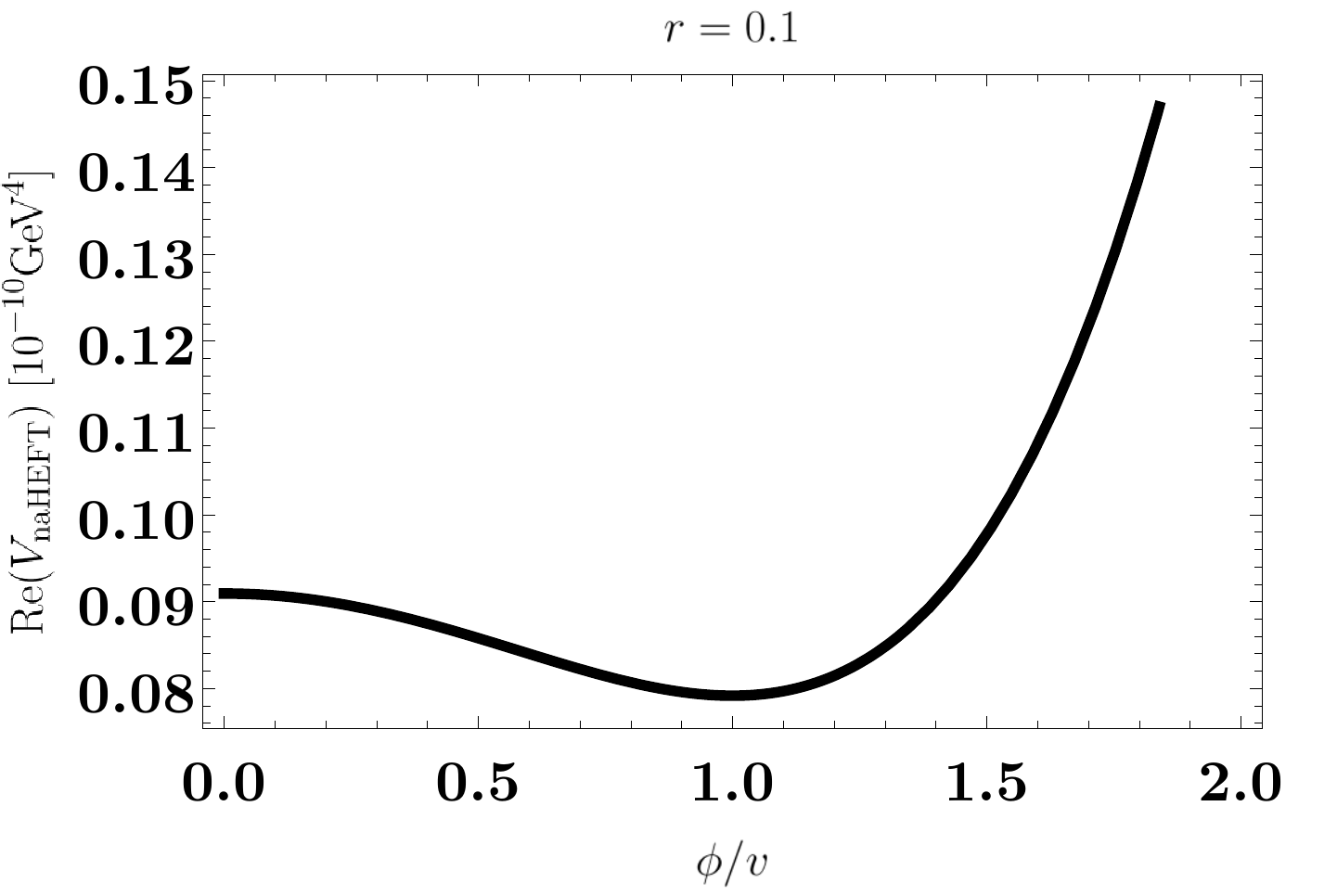}
	\includegraphics[width=7cm,clip]{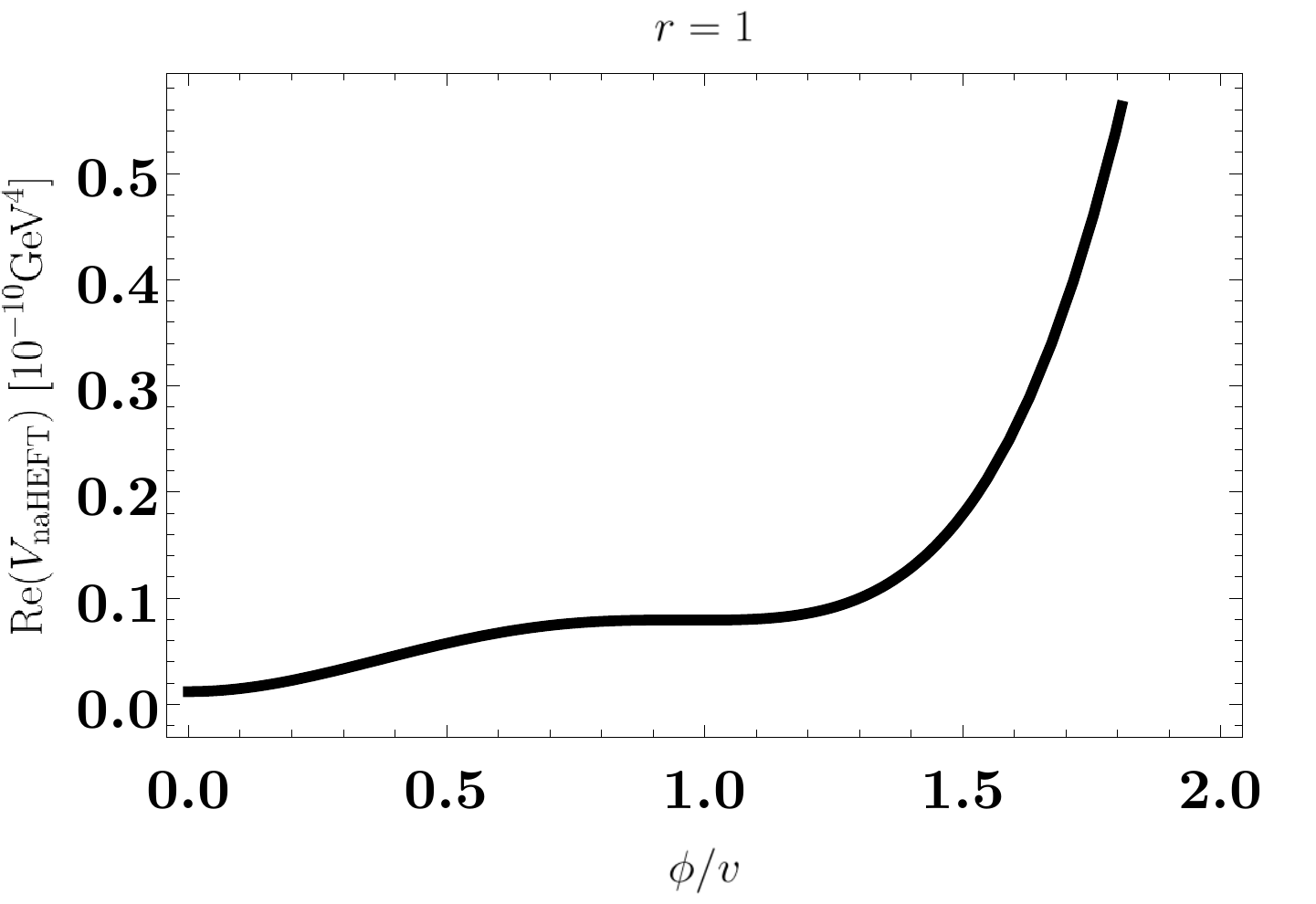}
	\caption{The real part of $V_{\rm{naHEFT}}$ (Eq.~(\ref{eq:Veff})). 
	We take $\Lambda=1\,\mbox{TeV}$ and $\kappa_0=1$.}
	\label{fig:Pottential}
\end{figure}
Let us consider the simple case where $\mathcal{K}(h)=0$ and 
$\mathcal{M}^2(h)$ is given as Eq.~(\ref{eq:simple-M}).
Using Eqs~(\ref{eq:dhat1}) and (\ref{eq:Mhsq}), we can eliminate $m^2$ and $\lambda$, and we find \cite{Anderson:1991zb}
\begin{align} 
V_{\rm{naHEFT}}
&\,=\,
\frac{M^2_h}{8v^2}\l(\phi^2-v^2\r)^2
\nn\\
&
+
\frac{\xi}{4}
\kappa_0
\biggl[
[\mathcal{M}^2(\phi)]^2\ln\frac{\mathcal{M}^2(\phi)}{\Lambda^2}
-\frac{3}{2}[\mathcal{M}^2(\phi)]^2
+2\,\Lambda^2\mathcal{M}^2(\phi)
\biggr]
\,,
\label{eq:Veff}
\end{align}
where
\begin{align}
\mathcal{M}^2(\phi)
\,=\,
\Lambda^2
+\frac{\kappa_{\rm{p}}}{2}
\l(\phi^2-v^2\r)
\,.
\end{align}
We here ignore the terms independent of $\phi$. 
We note that the scalar potential does not depend on the scale $\mu$ thanks to Eq.~(\ref{eq:simple-Meq}). 
In  \ref{fig:Pottential}, we show the typical structure of the Higgs potential (\ref{eq:Veff}).
We here fix $\Lambda=1\mbox{TeV}$ and $\kappa_0=1$.
We find that, if we take $|r|\simeq \mathcal{O}(1)$, $\phi=v$ is not ensured to be the global minimum. 
We will numerically estimate the vacuum stability bound later.

\section{Perturbative unitarity} 
\label{sec:unitarity}
In an EFT where the particle content is the same as the SM one,
the Higgs coupling deviation from the SM prediction causes the violation of perturbative unitarity in the scattering amplitudes. 
The energy scale of the perturbative unitarity violation can be regarded as the new physics 
scale at which unitarity is recovered by new particles and/or non-perturbative effects.
In this section we estimate the energy scale of the perturbative unitarity violation in the naHEFT framework, 
and discuss the relationship between the new physics scale and the Higgs coupling deviation.

We consider $S$-wave amplitudes for elastic $2\to 2$ scatterings of the longitudinally polarized $W$ and $Z$ bosons and the Higgs boson at high energies.
We compute the high-energy scattering amplitudes with the help of the equivalence theorem \cite{Cornwall:1974km,Lee:1977eg, Chanowitz:1985hj,Gounaris:1986cr,He:1993yd,He:1993qa}, which ensures the equality between the NG bosons and the longitudinal electroweak gauge bosons at the high energies. 
The relevant interactions for the scattering among the NG bosons and Higgs bosons can be read from the canonical Lagrangian, (\ref{eq:Lint}). 
From the unitarity argument on these amplitudes, we can obtain the upper bound on the scale of unitarity violation as a function of the coupling deviations in the Higgs-Gauge sector.{\footnote{The 
$2\to 2$ amplitudes generally do not provide the severest bound on the scale of unitarity violation. According to Refs.~\cite{Chang:2019vez, Falkowski:2019tft, Cohen:2021ucp}, the stronger bounds come from higher-multiplicity amplitudes.}}

In the high energy limit, the $S$-wave scattering matrix ($\mathcal{T}_0$) in the basis  
of $\ket{\pi^+\pi^-}$\,,
$\frac{1}{\sqrt{2}}\ket{\pi^3\pi^3}$\,,
$\frac{1}{\sqrt{2}}\ket{hh}$\,,
$\ket{h\pi^3}$
is
obtained as
\begin{align}
\mathcal{T}_0
&\,=\,
\frac{s}{32\pi v^2}
\l(
\begin{array}{cccc}
\tilde{A} & 
\sqrt{2}\tilde{A} &
\sqrt{2}\tilde{B} &
0
\\
\sqrt{2}\tilde{A} & 
0 &
\tilde{B} &
0
\\
\sqrt{2}\tilde{B} & 
\tilde{B} &
0 &
0
\\
0 & 
0 &
0 &
-\tilde{A}
\\
\end{array}
\r)
\nn\\
&
\qquad
\,-\,
\frac{M^2_h}{8\pi v^2}
\l(
\begin{array}{cccc}
A & \frac{1}{\sqrt{8}}B & \frac{1}{\sqrt{8}}C & 0\\
\frac{1}{\sqrt{8}}B & \frac{3}{4}D & \frac{1}{4}C & 0\\
\frac{1}{\sqrt{8}}C & \frac{1}{4}C & \frac{3}{4}E & 0\\
0 & 0 & 0 & \frac{1}{2}C
\end{array}
\r)+\mathcal{O}\l(\frac{M^2_h}{s}\r)
\,,
\label{eq:Tmat}
\end{align}
where $s$ denotes the center of mass energy squared and 
\begin{align}
\tilde{A}
&\,=\,
1-\kappa^2_V
\,,\\
\tilde{B}
&\,=\,
\kappa^2_V-\kappa_{VV}
\,,\\
A
&\,=\,
\kappa^2_V+(1+\kappa^2_V)\frac{M^2_W}{M^2_h}
\,,\\
B
&\,=\,
\kappa^2_V
\,,\\
C
&\,=\,
\kappa_V\l(3\kappa_3-2\kappa_V\r)
\,,\\
D
&\,=\,
\kappa^2_V\l(1+\frac{4M^2_Z}{3M^2_h}\r)
\,,\\
E
&\,=\,
\kappa_4
\,.
\end{align}
$M_W$ and $M_Z$ denote the mass of $W$ and $Z$ bosons, respectively. 
Here we compute the scattering amplitudes in the 't Hooft Feynman gauge. 
$\kappa$ parameters can be expressed in terms of the parameters in Eq.~(\ref{eq:BSM}). 
We summarize the detail of the computations in appendix \ref{app;amp}.

We estimate the perturbative unitarity bound by imposing 
\begin{align}
t_{\rm{max}}\,=\,\max_i
|\mbox{Re}(t_{i})|<t_0
\qquad \mbox{at}
~~
\sqrt{s}=\Lambda
\,,
\label{eq:unitairtybound}
\end{align}
where $t_i$ denotes the eigenvalues of $\mathcal{T}_0$ with $i$ being label of the eigenvalues.  
We impose this condition at the maximum energy scale of the EFT analysis, $\sqrt{s}=\Lambda$, which is regarded as the mass scale of the integrated new particles. 
The maximum value $t_0$ is conventionally taken to be 1/2 \cite{Gunion:1989we} or 1 \cite{Lee:1977eg}. In our analysis, we take $t_0=1/2$.

Before the numerical estimation of the perturbative unitarity bound, 
we report the useful analytic expressions.
We find that,
in the $M_{W,Z}\to 0$ limit, eigenvalues of the scattering matrix (\ref{eq:Tmat}) can be obtained analytically. 
Using the $\kappa$ notation defined in Eq.~(\ref{eq:Lint}), 
they are given by
\begin{align}
&t_1
\,=\,
\frac{s}{32\pi v^2}(\kappa^2_V-1)
-\frac{M^2_h}{16\pi v^2}\kappa^2_V
\,,
\label{eq:t1-kappa}
\\
&t_2
\,=\,
\frac{s}{32\pi v^2}(1-\kappa^2_V)
-\frac{M^2_h}{16\pi v^2}
\frac{3\kappa_4+5\kappa^2_V+4\sqrt{K(s)}}{4}
\,,\\
&t_3
\,=\,
\frac{s}{32\pi v^2}(1-\kappa^2_V)
-\frac{M^2_h}{16\pi v^2}
\frac{3\kappa_4+5\kappa^2_V-4\sqrt{K(s)}}{4}
\,,\\
&
t_4
\,=\,
\frac{s}{32\pi v^2}(\kappa^2_V-1)
-\frac{M^2_h}{16\pi v^2}
\kappa_V(3\kappa_3-2\kappa_V)
\,,
\label{eq:t4-kappa}
\end{align}
where
\begin{align}
&
K(s)
\,=\,
\Delta_1\frac{s^2}{M^4_h}
-\Delta_2\frac{s}{M^2_h}
+
\frac{1}{16}\biggl[
9\kappa^2_4
+\kappa^2_V
(73\kappa^2_V-144\kappa_3\kappa_V+108\kappa^2_3-30\kappa_4)
\biggr]
\,,\\
&
\Delta_1
\,=\,
(\kappa^2_V-1)^2
-\frac{3}{2}(\kappa^2_V-1)(\kappa_{VV}-1)
+\frac{3}{4}(\kappa_{VV}-1)^2
\,,\\
&
\Delta_2
\,=\,
\frac{9}{2}\kappa_3\kappa_V(\kappa^2_V-\kappa_{VV})
+\frac{3}{4}\kappa_4(\kappa^2_V-1)
-\frac{1}{4}\kappa^2_V(17\kappa^2_V-12\kappa_{VV}-5)
\,.
\end{align}
We note that  $\Delta_1=\Delta_2=0$ and $K(s)=1$ in the SM ($\kappa_V=\kappa_{VV}=\kappa_3=\kappa_4=1$).
Using Eqs.~(\ref{eq:kappa3-naHEFT}) and (\ref{eq:kappa4-naHEFT}) and setting $s=\Lambda^2$, we obtain
\begin{align}
&
t_1
\,=\,
\frac{\Lambda^2}{16\pi v^2}
\Delta \kappa_V
\,,
\label{eq:t1app}
\\
&
t_2
\,=\,
-
\frac{3\Lambda^2}{64\pi v^2}
\biggl[
\Delta\kappa_{VV}
+8\,\xi\,\kappa_0\,r^3
\l(1-\frac{r}{6}\r)
\frac{\Lambda^2}{v^2}
\biggr]
\,,\\
&
t_3
\,=\,
-
\frac{\Lambda^2}{8\pi v^2}
\biggl[
\Delta\kappa_V
-\frac{3}{8}\Delta\kappa_{VV}
+3\,\xi\,\kappa_0\,r^3
\l(1-\frac{r}{2}\r)
\frac{\Lambda^2}{v^2}
\biggr]
\,,\\
&
t_4
\,=\,
-
\frac{\Lambda^2}{16\pi v^2}
\biggl(
\Delta\kappa_V
+4\,\xi\,\kappa_0\,r^3\,
\frac{\Lambda^2}{v^2}
\biggr)
\,,
\label{eq:t4app}
\end{align}
where we ignore the $\mathcal{O}(\xi^2,M^2_h/\Lambda^2, (\Delta\kappa_V)^2, (\Delta\kappa_{VV})^2)$ corrections.
We find that, if either $\Delta\kappa_V$, $\Delta\kappa_{VV}$, or $r$ is nonzero, 
the eigenvalues grow as the cutoff scale, $\Lambda$.
Therefore, imposing the unitarity bound in Eq.~(\ref{eq:unitairtybound}), we can obtain the relationship between the cutoff scale $\Lambda$ and the Higgs coupling deviation factors.
We will perform the numerical analysis below.

\section{Numerical analysis}
\label{sec:numeric}
In this section, we numerically estimate vacuum stability and perturbative unitarity bound in the naHEFT.  
In the estimation of the perturbative unitarity bound, we use the expressions reported in appendix \ref{app;amp} and diagonalize the scattering matrix numerically. For concreteness,
we consider the following two cases; i) the simple case in which non-decoupling effect only appear in the Higgs potential,
 and ii) the case for the scalar extension which has been discussed in section \ref{sec:model-scalar}.
\subsection{The simple case}
\label{sec:simp-case}
\begin{figure}[t]
	\centering
	\includegraphics[width=7cm,clip]{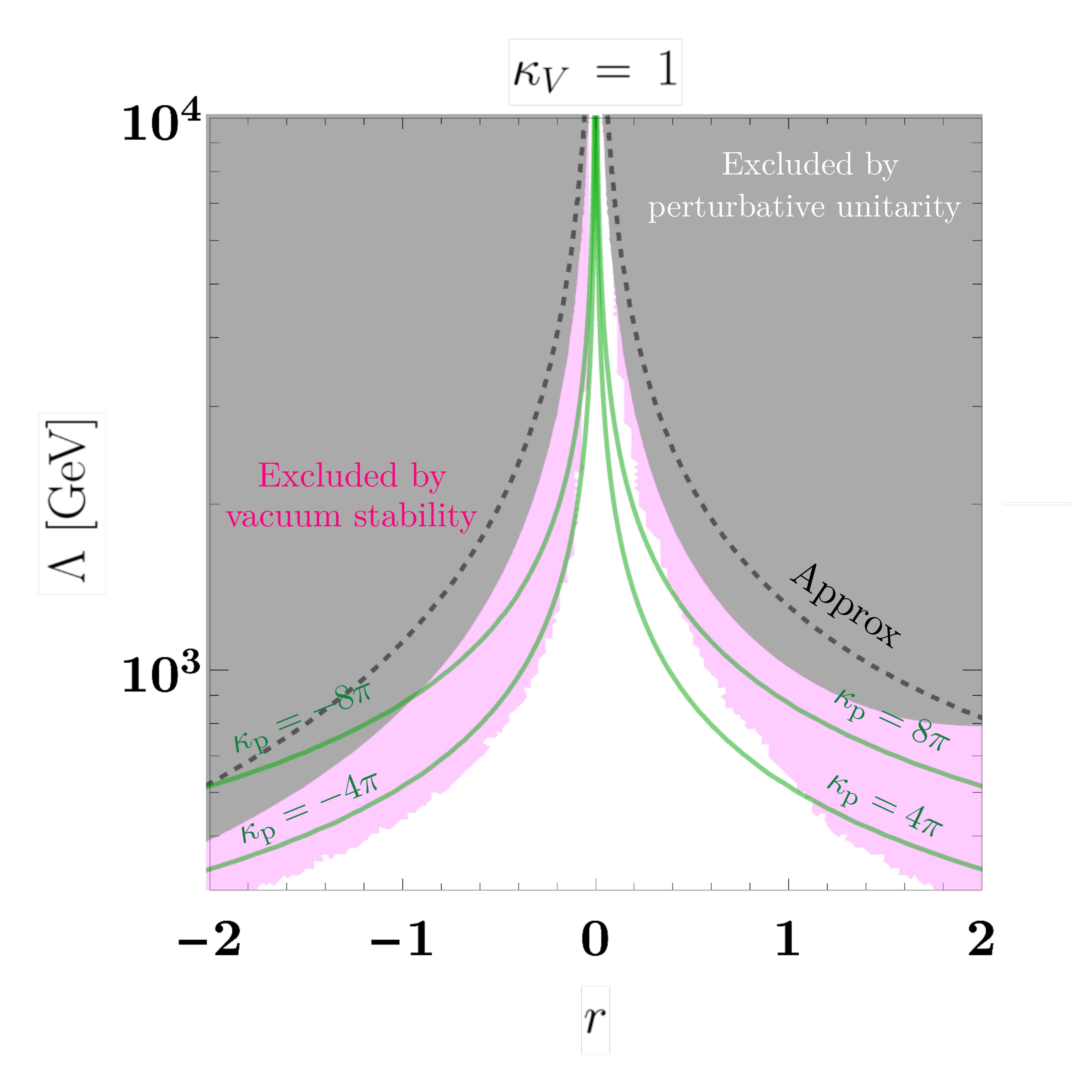}
	\includegraphics[width=7cm,clip]{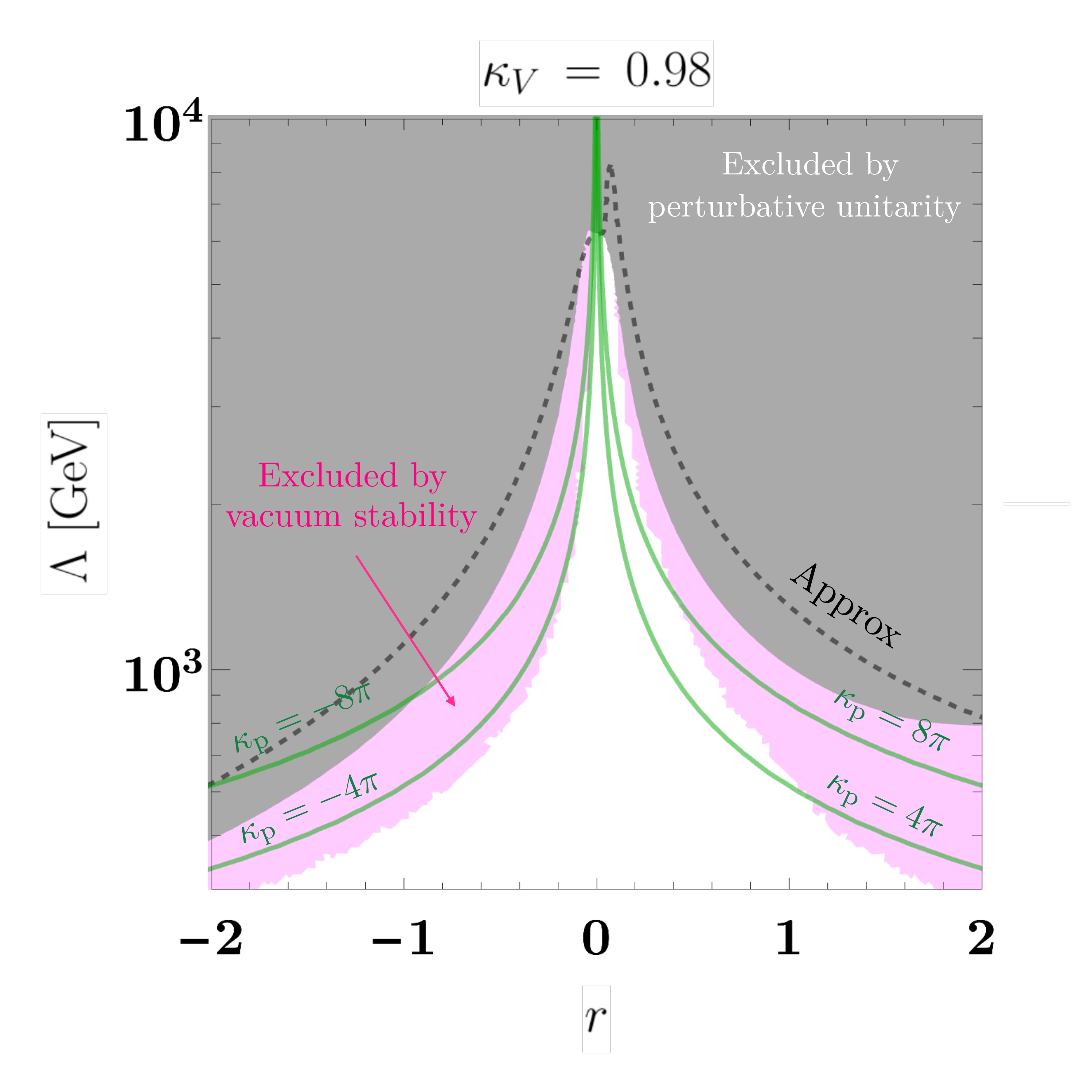}
	\caption{
	Perturbative unitarity bound (Gray) and vacuum stability bound (Magenta) for the simple case discussed in section \ref{sec:simp-case}. We take $\kappa_{VV}=1$. Dotted line is obtained by using Eqs.~(\ref{eq:t1app})-(\ref{eq:t4app}).}
	\label{fig:LQT-simp-Lam-r}
\end{figure}
We first consider the simple case where
\begin{itemize}
\item 
$\mathcal{M}^2(h)=\Lambda^2+\kappa_{\rm{p}}\l(|\Phi|^2-\frac{v^2}{2}\r)$
\,,
\item
$\mathcal{K}(h)=0$\,,
\item
$\mathcal{F}(h)$ is independent of $\Lambda^2$ and $\kappa_{\rm{p}}$\,,
\item
$\kappa_0=1$.
\end{itemize}
In this case, the Higgs potential is estimated as Eq.~(\ref{eq:Veff})
and the Higgs couplings relevant to the unitarity bound are obtained as
\begin{align}
\kappa_V
&\,=\,
1+\frac{\xi}{2}f_1
\,,
\label{eq:kappaV-simp}\\
\kappa_{VV}
&\,=\,
1+\frac{\xi}{2}f_2
\,,\\
\kappa_{3}
&\,=\,
1+\frac{4}{3}\xi \frac{\Lambda^4}{v^2 M^2_h} r^3
\,,
\label{eq:kappa3-simp}
\\
\kappa_{4}
&\,=\,
1+\frac{16}{3}\xi \frac{\Lambda^4}{v^2 M^2_h} \frac{r^3(r-3)}{2}
\,,
\label{eq:kappa4-simp}
\end{align}
where $r=\frac{\frac{\kappa_{\rm{p}}}{2}v^2}{\Lambda^2}$. 
For $|r|\sim \mathcal{O}(1)$, the significant non-decoupling effects appear in $\kappa_3$ and $\kappa_4$.
There are four parameters to be constrained by vacuum stability and perturbative unitarity;
\begin{align}
f_1\,,\quad
f_2\,,\quad
\Lambda\,,\quad
r\,.
\end{align}
In the following analysis, we convert the parameter set into 
\begin{align}
\kappa_V\,,\quad
\kappa_{VV}\,,\quad
\Lambda\,,\quad
r\,,
\end{align}
or 
\begin{align}
\kappa_V\,,\quad
\kappa_{VV}\,,\quad
\Lambda\,,\quad
\kappa_3\,,
\end{align}
using Eqs.~(\ref{eq:kappaV-simp})-(\ref{eq:kappa4-simp}).

\begin{figure}[t]
	\centering
	\includegraphics[width=7cm,clip]{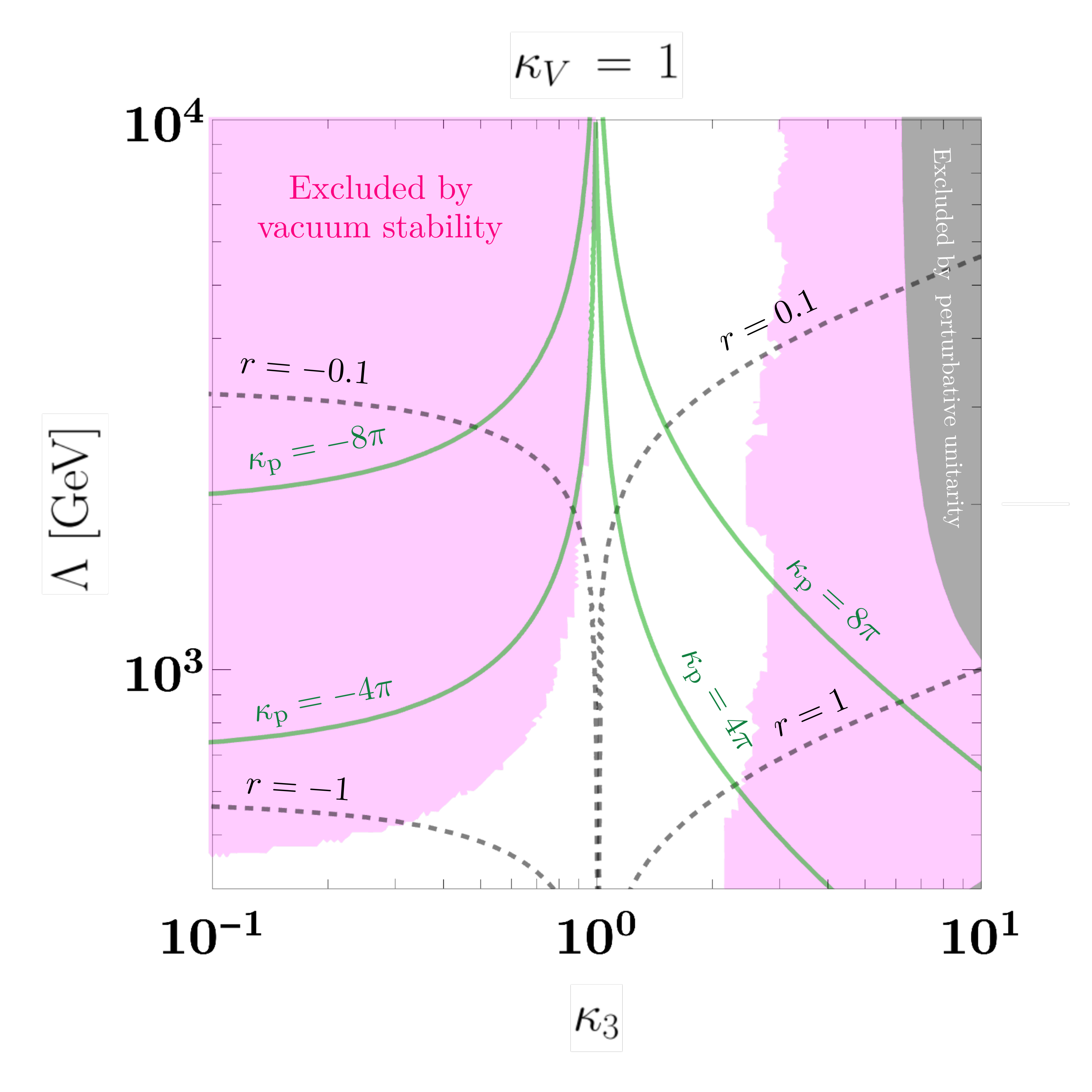}
	\includegraphics[width=7cm,clip]{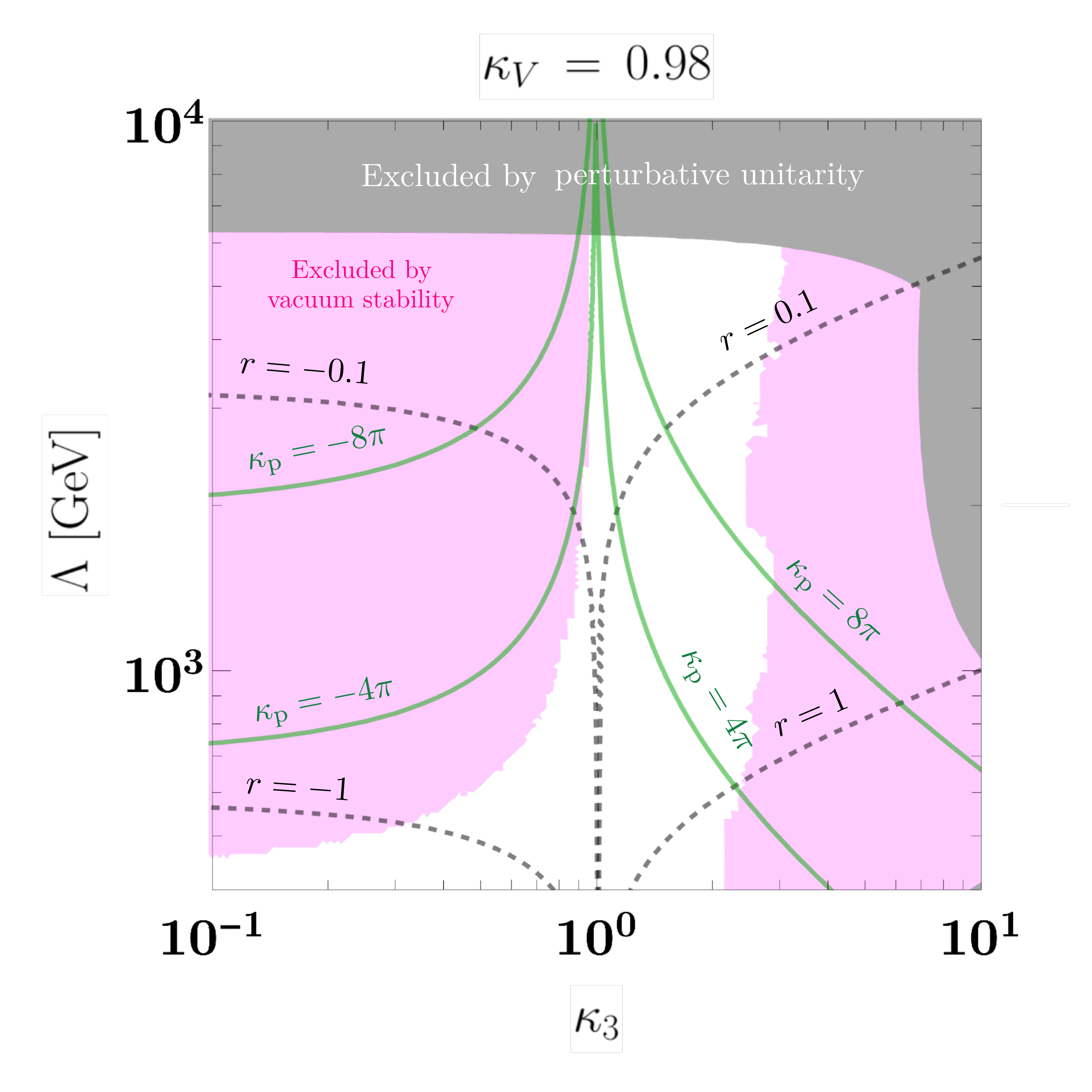}
	\caption{Similar plots to figure \ref{fig:LQT-simp-Lam-r}. We employ $\kappa_3$ as an input parameter instead of $r$. Black dotted lines are contours of $r=-1,-0.1,0.1$, and $1$.}
	\label{fig:LQT-simp-Lam-k3}
\end{figure}
In figure \ref{fig:LQT-simp-Lam-r}, we show the vacuum stability and perturbative unitarity bounds in the $(\Lambda, r)$ plane with $\kappa_V$ and $\kappa_{VV}$ being fixed.
We take $\kappa_V=1$ and $0.98$ in left and right figures, and fix $\kappa_{VV}=1$ in both figures.
The magenta and gray regions are excluded by vacuum stability and perturbative unitarity, respectively. 
The black dotted line is the contour of $t_{\rm{max}}=1/2$ obtained by using analytic expressions in Eqs.~(\ref{eq:t1app})-(\ref{eq:t4app}). 
We obtain the upper bound on the cutoff scale if either $\kappa_V\neq 1$ or $r\neq 0$.
When $r\simeq 0$ and $\kappa_V\neq 1$, the perturbative unitarity bound gives the stringent constraint, which is roughly given as $\Lambda^2\lesssim \frac{8\pi v^2}{1-\kappa^2_V}$. For the other parameter region, the vacuum stability bound is stronger than the perturbative unitarity bound. 

The green lines are contours of $|\kappa_{\rm{p}}|=4\pi$ and $8\pi$.
The parameter region where $|\kappa_{\rm{p}}|\gtrsim 4\pi, 8\pi$ may be dangerous regions because the ``portal'' coupling between the integrated new particle and the Higgs field becomes too large to keep the perturbativity.

We map the vacuum stability and unitarity bounds from the $(\Lambda, r)$ plane to $(\Lambda, \kappa_3)$ plane in figure \ref{fig:LQT-simp-Lam-k3}.
The black dotted lines are contours of $|r|=0.1$ and $1$.
We find that the vacuum stability bound excludes $0.9\lesssim \kappa_3$ and $\kappa_3\gtrsim 2$ for $\Lambda \gtrsim 1\,\mbox{TeV}$.
For $\Lambda \lesssim 1\,\mbox{TeV}$, the excluded region for $\kappa_3<1$ is sensitive to $\Lambda$.

\subsection{Extended scalar model}
\begin{figure}[t]
	\centering
	\includegraphics[width=7cm,clip]{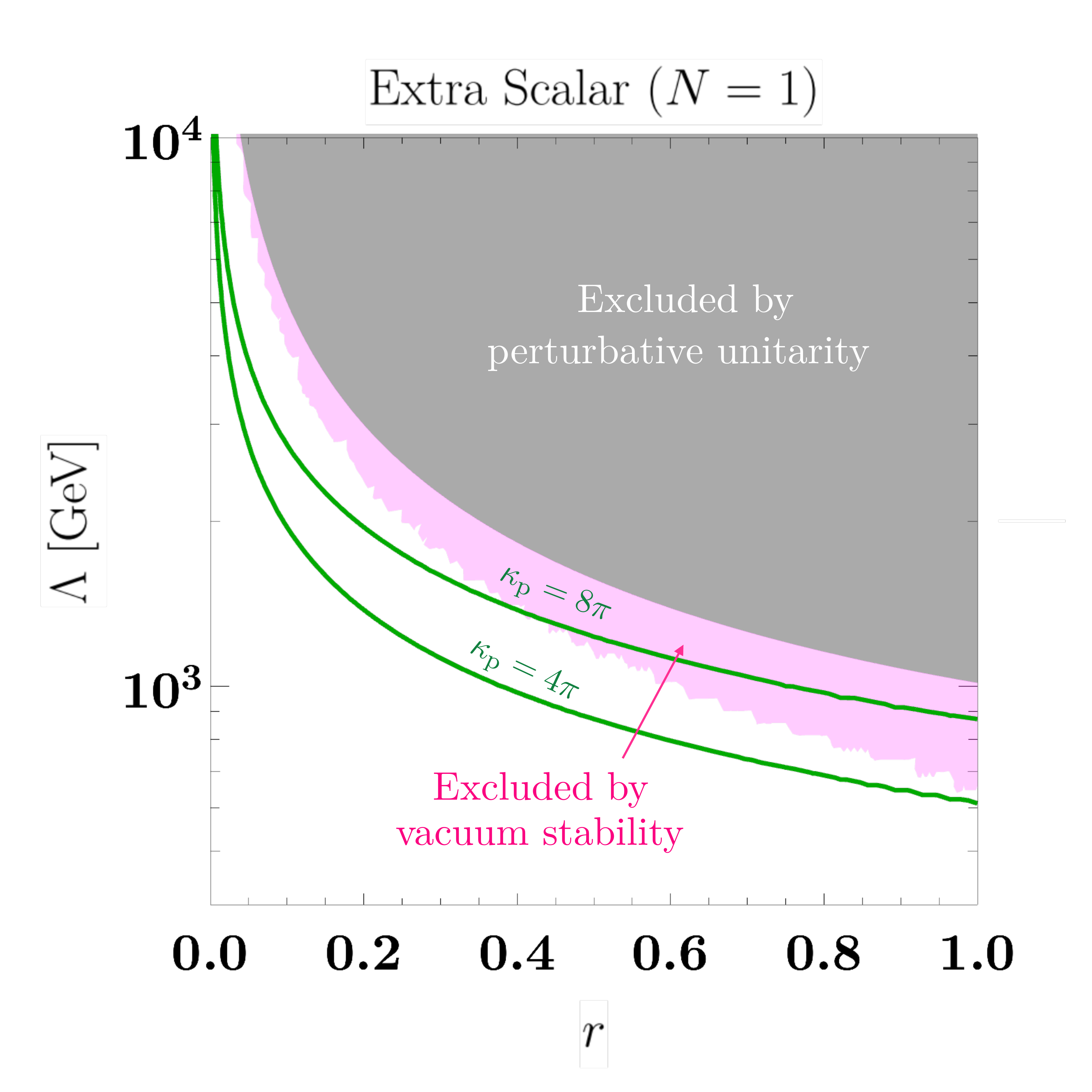}
	\includegraphics[width=7cm,clip]{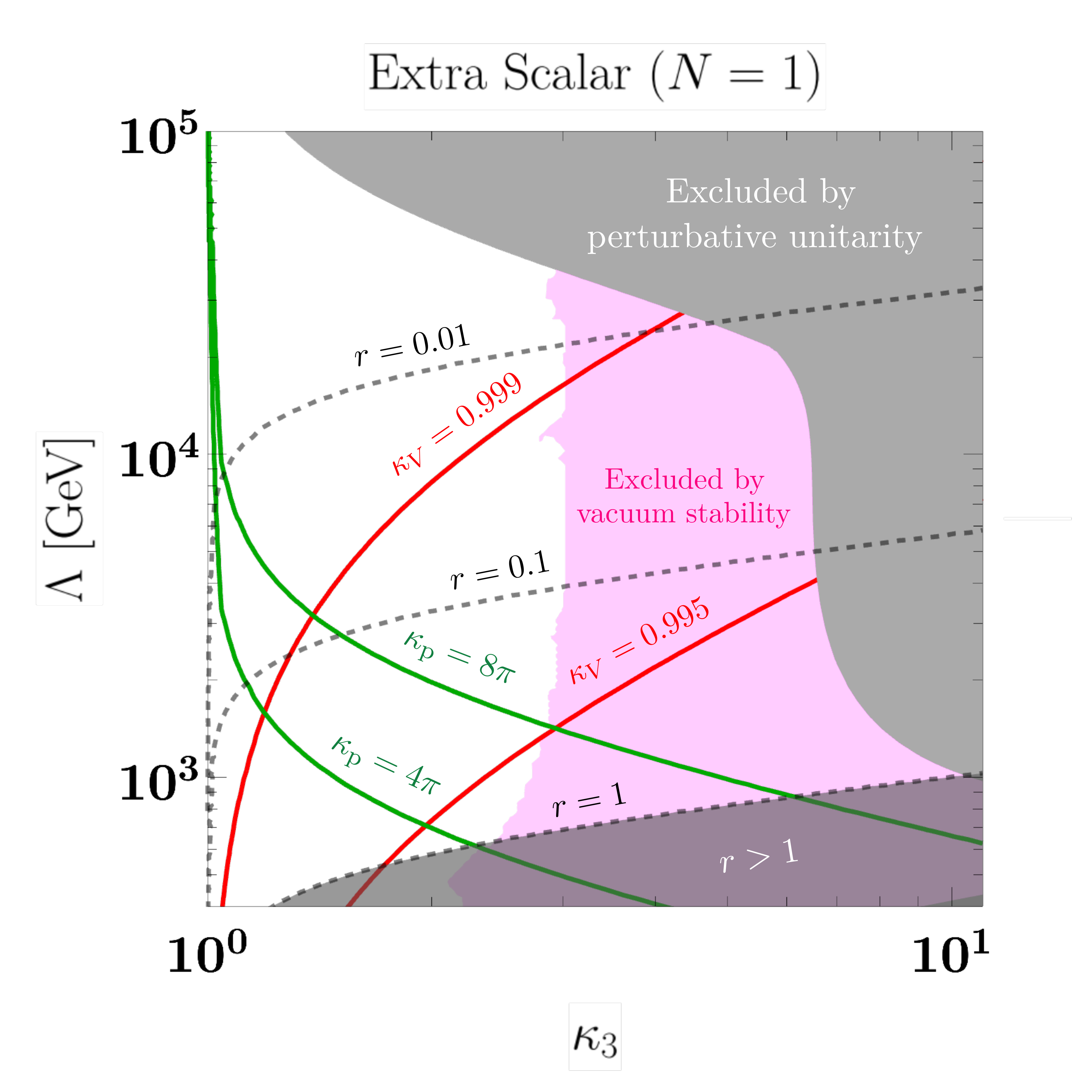}
	\\
        \includegraphics[width=7cm,clip]{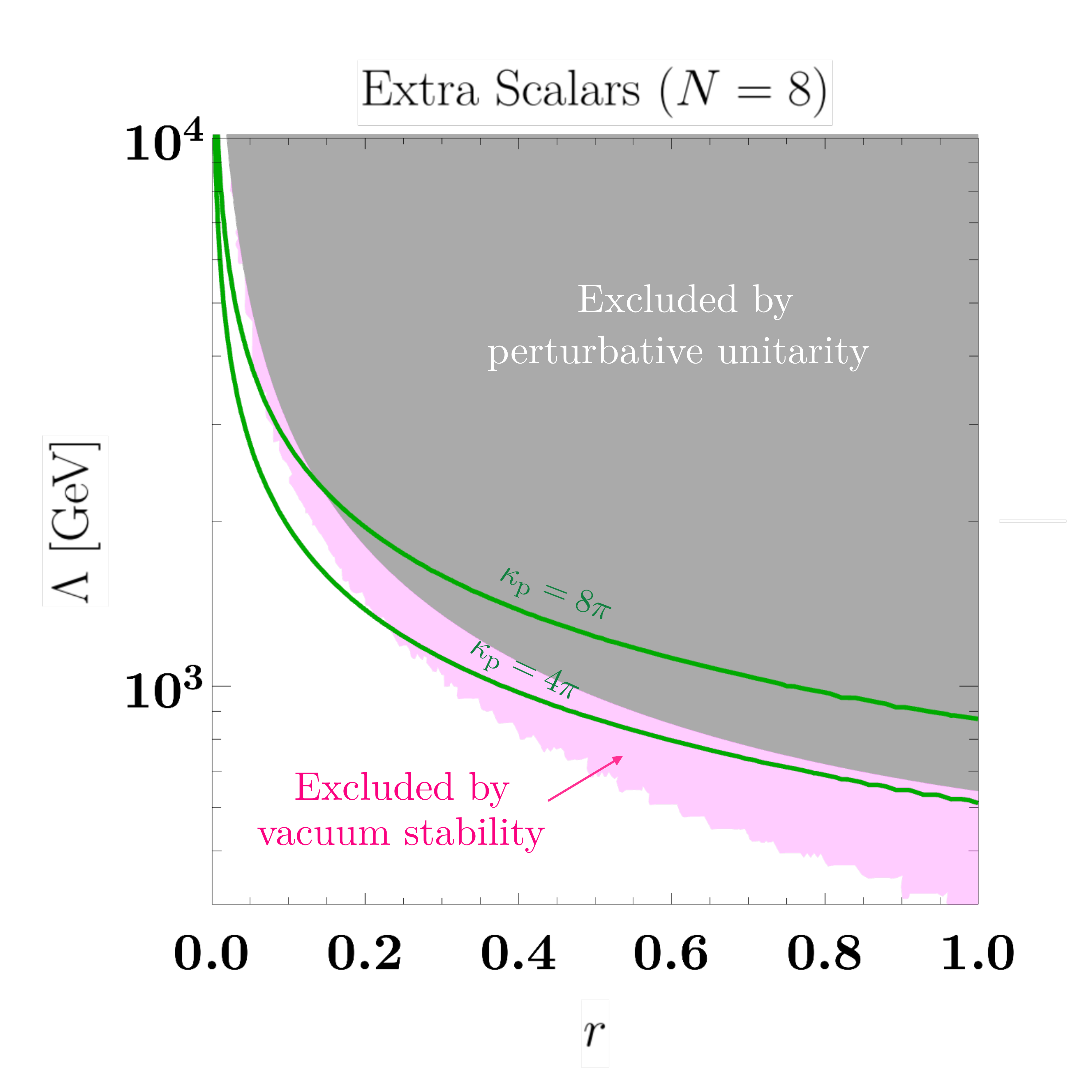}
	\includegraphics[width=7cm,clip]{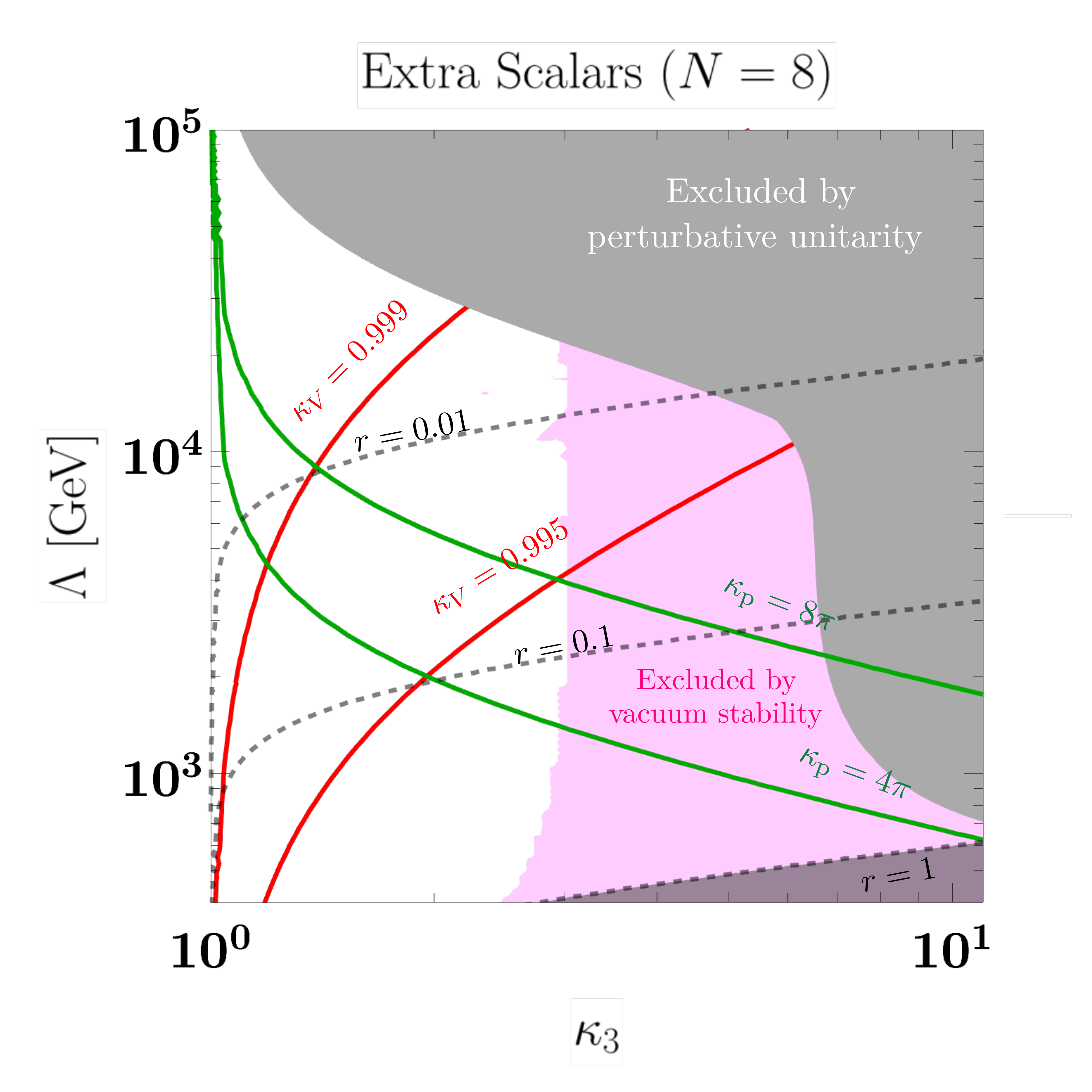}
	\caption{Perturbative unitarity bound (Gray) and vacuum stability bound (Magenta) for the case discussed in section \ref{sec:model-scalar}. We take $N=1$ (Top) and $8$ (Bottom).}
	\label{fig:LQT-scalar-Lam-r}
\end{figure}
We next consider the case discussed in section \ref{sec:model-scalar}.
New physics effects in this case are parameterized by the polynomials, (\ref{eq:HSM-M})-(\ref{eq:HSM-zeros}). 
The relevant Higgs couplings are obtained as Eq.~(\ref{eq:kappaV-scalar})-(\ref{eq:kappa4-scalar}).
In contrast with the previous case, $\kappa_V$ and $\kappa_{VV}$ are determined by $\Lambda$ and $r$. Therefore, the free parameters are
\begin{align}
\Lambda\,,\qquad
r\,,\qquad
N\,.
\end{align}

Figure \ref{fig:LQT-scalar-Lam-r} shows vacuum stability and perturbative unitarity bounds for $N=1$ and $8$ cases. 
Let us first focus on the left figures in which we plot these constraints on the $(\Lambda,r)$ plane.
The color notation is the same as figure \ref{fig:LQT-simp-Lam-r}.
We note that the upper bound on $\Lambda$ is obtained for $r\neq 0$,
and the vacuum stability bound is stronger than the perturbative unitarity bound.
We cannot obtain the upper bound on $\Lambda$ when $r=0$ because the Higgs couplings do not deviate from the SM prediction if $r=0$.

In the right figures, 
we show the excluded region in the $(\Lambda, \kappa_3)$ plane. We here use Eq.~(\ref{eq:kappa3-scalar}) to convert the parameter set $(\Lambda,r)$ into $(\Lambda,\kappa_3)$. 
Black dotted lines are contours of $r=0.01,0.1$, and $1$. 
The dark gray shaded region in the bottom of the figure corresponds to the region where $r > 1$, which we do not consider in our analysis. 
Red lines are contours of $\kappa_V=0.999$ and $0.995$.
We find that the vacuum stability bound excludes $\kappa_3\gtrsim 3$. 
For $\kappa_3\lesssim 3$, the perturbative unitarity constraints put the upper bound on $\Lambda$, which is roughly given as $\Lambda^2\lesssim \frac{8\pi v^2}{1-\kappa^2_V}$. 

\section{New No-Lose Theorem}
\label{sec:nolose}
We have obtained the scale of new physics as a function of the Higgs coupling deviations by imposing vacuum stability and perturbative unitarity in the naHEFT. 
Our finding means that measurements of the Higgs coupling deviation (induced by the non-decoupling effects) directly point to the energy scale where new physics must appear.  
Therefore, if the deviation is observed at future colliders, it would provide a target energy scale for future colliders designed to observe the new particles and/or non-perturbative effects directly.

Our argument follows the famous classic argument for the new physics scale by Lee, Quigg, and Thacker (LQT) in 1977 \cite{Lee:1977eg,Lee:1977yc}. 
They have estimated the upper bound on the new physics scale using perturbative unitarity in elastic scatterings of longitudinally polarized $W$ and $Z$ bosons in an electroweak theory without the 125\,GeV Higgs boson.  
They have found that the new physics scale $\Lambda$ must satisfy
\begin{align}
\Lambda^2
\,\lesssim \,
8\pi v^2
\simeq 
(1.2\,\mbox{TeV})^2
\,.
\end{align}
They have also discussed the perturbative unitarity in an electroweak theory with a Higgs boson, and they have shown that the Higgs boson mass must be lower than $(\frac{16\pi v^2}{3})^{1/2}\simeq 1$\,TeV. 
Their argument has been one of the most important motivations for the energy scale of the LHC, which is refereed to as the ``no-lose theorem''.

After the 125\,GeV Higgs discovery, the perturbative unitarity argument was applied to an electroweak theory with the 125\,GeV Higgs boson, $h$. If the Higgs coupling deviates from the SM prediction without new particles, the perturbative unitarity violation occurs at some energy scales even with the presence of the 125\,GeV Higgs boson. If one focuses on elastic scatterings of longitudinally polarized $W$ and $Z$ bosons, 
the energy scale of the perturbative unitarity violation is roughly estimated as \cite{Nagai:2014cua}
\begin{align}
\Lambda^2
\,\lesssim \,
\frac{8\pi v^2}{|1-\kappa^2_V|}
\,,
\end{align} 
where $\kappa_V$ denotes the Higgs-Gauge-Gauge coupling with $\kappa_V=1$ in the SM.
This relation means that, if we will observe the Higgs coupling deviation ($\kappa_V\neq 1$),
there should be new particle(s) below $\l(\frac{8\pi v^2}{|1-\kappa^2_V|}\r)^{\frac{1}{2}}$ to recover the perturbative unitarity.
This is the modern version of LQT's ``no-lose theorem'', which can give a motivation of future collider experiments to probe the new particles.
In renormalizable BSMs, the perturbative unitarity violation scale can be regarded as the mass scale of new particles. See Refs.~\cite{Kanemura:1993hm,Akeroyd:2000wc,Kanemura:2015ska} for the complete studies of the two-Higgs doublet model, for example.

In this paper, we extend above unitarity arguments with including the non-decoupling effects to the Higgs self-couplings. 
We also estimate the vacuum stability bound.
In our EFT formalism, the new physics scale (cutoff scale) $\Lambda$ is defined by the internal parameter in the Higgs potential as Eq.~(\ref{eq:defLam}). 
As results, we found that even if $\kappa_V=1$ 
the non-trivial constraints on $\Lambda$ can be obtained when $r\neq 0$.
See figure \ref{fig:LQT-simp-Lam-r} for instance.
We emphasize that this bound cannot be obtained using the conventional EFT, where the new physics effects are described by the set of higher dimensional operators with finite number truncation. 
We refer to our findings as the ``{\it{new}} no-lose theorem'' inspired by the successful history of the related arguments.

\section{Conclusions}
\label{sec:summary}
We constructed the new effective field theory (the naHEFT) based on the assumption that i) the new physics scale is higher than the electroweak scale and ii) new physics effects enter into the low-energy theory via the quantum effects. The most remarkable feature of the naHEFT is to be able to express non-decoupling quantum effects from the new physics, which cannot be parameterized by the conventional linear EFTs.  
We confirmed the validity of the naHEFT by comparing with the results obtained in the UV completed model.

In section \ref{sec:vacuum stability}, we discussed the vacuum stability bound in the naHEFT framework.
We imposed that the electroweak vacuum is the global minimum and we obtained the upper bound on the non-decouplingness of new physics from the vacuum stability argument. 

We next discussed the perturbative unitarity bound in section \ref{sec:unitarity}.
We calculated the high-energy $2\to2$ $S$-wave scattering amplitudes among the NG bosons and the Higgs bosons, and imposed that the largest eigenvalue of the scattering matrix does not exceed the unitarity bound. 
We then obtained the upper bound on the new physics scale as a function of the Higgs coupling deviation and the non-decouplingness. 

Combining the vacuum stability and perturbative unitarity bounds, 
we numerically evaluated the scale of new physics as a function of the Higgs coupling deviations.
We confirmed that the new physics scale can be obtained if we observe the Higgs coupling deviation via non-decoupling effects at future collider experiments.
We referred to the relation between the Higgs coupling deviation and the new physics scale as the ``new no-lose theorem'', which is inspired by the successful arguments by Lee, Quigg, and Thacker in 1977.

Possible outlook of the naHEFT is to investigate the electroweak phase transition (EWPT) in early universe. It has been known that, in some concrete new physics models, the non-decoupling effect in the Higgs potential plays an important role for realizing the strongly first order phase transition,
which yields the gravitational wave with the specific spectrum \cite{Kakizaki:2015wua, Hashino:2016rvx}. 
Since the naHEFT can parameterize the non-decoupling effect systematically, we believe that the naHEFT is the best EFT framework for investigating the EWPT in a model-independent way.  
This is performed elsewhere \cite{SKandRN}.

We finally emphasize that, in addition to the direct search of the new particles, 
the Higgs coupling measurements tell us important clues for investigating physics beyond the SM. Future precision measurements of the Higgs property at future collider such as the LHC, ILC experiments and so on will be able to pin down the direction of the new physics beyond the SM.

\section*{Acknowledgments}
This work of S.K. was supported, in part, by the Grant-in-Aid on Innovative Areas, the Ministry of Education, Culture, Sports, Science and Technology, No. 16H06492 and by the JSPS KAKENHI Grant No. 20H00160. 
The work of R.N. was supported by JSPS KAKENHI Grant Numbers JP19K14701 and JP21J01070. 
\appendix
\section{$S$-wave amplitudes}
\label{app;amp}
\paragraph{Lorentz invariant amplitudes}
We first compute the scattering amplitudes.
From Eq.~(\ref{eq:Lint}), we find
\begin{align}
&\mathcal{A}(\pi^+\pi^-\to\pi^+\pi^-)
\,=\,
-\frac{u}{v^2}
-\frac{\kappa^2_V}{v^2}
\l(\frac{s^2}{s-M^2_h}+\frac{t^2}{t-M^2_h}\r)
\,,\\
&\mathcal{A}(\pi^3\pi^3\to\pi^3\pi^3)
\,=\,
-\frac{\kappa^2_V}{v^2}
\l(
\frac{s^2}{s-M^2_h}
+
\frac{t^2}{t-M^2_h}
+
\frac{u^2}{u-M^2_h}
\r)
\,,\\
&\mathcal{A}(hh\to hh)
\,=\,
-\frac{3M^2_h}{v^2}
\biggl[
\kappa_4+3\kappa^2_3
\l(\frac{M^2_h}{s-M^2_h}
+\frac{M^2_h}{t-M^2_h}
+\frac{M^2_h}{u-M^2_h}\r)
\biggr]
\,,\\
&\mathcal{A}(\pi^+\pi^-\to\pi^3\pi^3)
\,=\,
\frac{s}{v^2}
-\frac{\kappa^2_V}{v^2}
\frac{s^2}{s-M^2_h}
\,,\\
&\mathcal{A}(\pi^+\pi^-\to hh)
\,=\,
-\frac{s}{v^2}(\kappa_{VV}-\kappa^2_V)
\nn\\
&
\qquad \qquad 
-\frac{M^2_h}{v^2}\kappa_V
\biggl[
3\kappa_3-2\kappa_V
+3\kappa_3\frac{M^2_h}{s-M^2_h}
+\kappa_V\frac{M^2_h}{t-M^2_W}
+\kappa_V\frac{M^2_h}{u-M^2_W}
\biggr]
\,,\\
&\mathcal{A}(\pi^3\pi^3\to\pi^3\pi^3)
\,=\,
\frac{s}{v^2}
-\frac{\kappa^2_V}{v^2}
\frac{s^2}{s-M^2_h}
\,,\\
&\mathcal{A}(\pi^+\pi^-\to hh)
\,=\,
-\frac{s}{v^2}(\kappa_{VV}-\kappa^2_V)
\nn\\
&
\qquad \qquad 
-\frac{M^2_h}{v^2}\kappa_V
\biggl[
3\kappa_3-2\kappa_V
+3\kappa_3\frac{M^2_h}{s-M^2_h}
+\kappa_V\frac{M^2_h}{t-M^2_W}
+\kappa_V\frac{M^2_h}{u-M^2_W}
\biggr]
\,,\\
&\mathcal{A}(\pi^3\pi^3\to hh)
\,=\,
-\frac{s}{v^2}(\kappa_{VV}-\kappa^2_V)
\nn\\
&
\qquad \qquad
-\frac{M^2_h}{v^2}\kappa_V
\biggl[
3\kappa_3-2\kappa_V
+3\kappa_3\frac{M^2_h}{s-M^2_h}
+\kappa_V\frac{M^2_h}{t-M^2_Z}
+\kappa_V\frac{M^2_h}{u-M^2_Z}
\biggr]
\,,\\
&\mathcal{A}(h\pi^3\to h\pi^3)
\,=\,
-\frac{t}{v^2}(\kappa_{VV}-\kappa^2_V)
\nn\\
&
\qquad \qquad 
-\frac{M^2_h}{v^2}\kappa_V
\biggl[
3\kappa_3-2\kappa_V
+3\kappa_3\frac{M^2_h}{t-M^2_h}
+\kappa_V\frac{M^2_h}{s-M^2_Z}
+\kappa_V\frac{M^2_h}{u-M^2_Z}
\biggr]
\,,
\end{align}
where we use the 't Hooft Feynman gauge.
\paragraph{$S$ wave amplitudes}
We next estimate $S$ wave amplitudes defined by
\begin{align}
a_0(i)
&
\,=\,
\frac{1}{32\pi} \int^1_{-1} d\cos\theta\,
\mathcal{A}(i)
\,,
\end{align}
with $\theta$ being the scattering angle for the $2\to 2$ process.
Performing the scattering angle integrations, we obtain
\begin{align}
a_0&(\pi^+\pi^-\to \pi^+\pi^-)
\,=\,
\frac{s}{32\pi v^2}(1-\kappa^2_V)
\nn\\
&
-\frac{M^2_h}{16\pi v^2}
\kappa^2_V
\biggl[
2
+\frac{2M^2_W}{M^2_h}\frac{1+\kappa^2_V}{\kappa^2_V}
+\frac{M^2_h}{s-M^2_h}
-\frac{M^2_h}{s-4M^2_W}
\ln\l(
1+\frac{s-4M^2_W}{M^2_h}
\r)
\biggr]
\,,
\\
a_0&(\pi^3\pi^3\to \pi^3\pi^3)
\,=\,
-\frac{M^2_h}{16\pi v^2}\kappa^2_V
\biggl[
3+\frac{M^2_h}{s-M^2_h}+\frac{4M^2_Z}{M^2_h}
-\frac{2M^2_h}{s-4M^2_Z}
\ln\l(
1+\frac{s-4M^2_Z}{M^2_h}
\r)
\biggr]
\,,\\
a_0&(hh\to hh)
\,=\,
-\frac{M^2_h}{16\pi v^2}
\biggl[
3\kappa_4
+9\kappa^2_3 \frac{M^2_h}{s-M^2_h}
-18\kappa^2_3 \frac{M^2_h}{s-4M^2_h}
\ln\l(\frac{s}{M^2_h}-3\r)
\biggr]
\,,
\label{eq:a0hh2hh}
\\
a_0&(\pi^+\pi^- \to \pi^3\pi^3 )
\,=\,
\frac{s}{16\pi v^2}(1-\kappa^2_V)
-\frac{M^2_h}{16\pi v^2}\kappa^2_V
\l(1+\frac{M^2_h}{s-M^2_h}\r)
\,,\\
a_0&(\pi^+\pi^- \to hh )
\,=\,
\frac{s}{16\pi v^2}
(\kappa^2_V-\kappa_{VV})
\nn\\
&
-\frac{M^2_h}{16v^2}\kappa_V
\biggl[
3\kappa_3-2\kappa_V
+3\kappa_3\frac{M^2_h}{s-M^2_h}
\nn\\
&
+\kappa_V
\frac{4M^2_h}{\sqrt{(s-4M^2_W)(s-4M^2_h)}}
\ln\l(\frac{s-2M^2_h-\sqrt{(s-4M^2_W)(s-4M^2_h)}}
{2M^2_h\sqrt{1-\frac{4M^2_W}{M^2_h}+\frac{sM^2_W}{M^4_h}}}\r)
\biggr]
\,,\\
a_0&(\pi^3\pi^3 \to hh )
\,=\,
\frac{s}{16\pi v^2}
(\kappa^2_V-\kappa_{VV})
\nn\\
&
-\frac{M^2_h}{16v^2}\kappa_V
\biggl[
3\kappa_3-2\kappa_V
+3\kappa_3\frac{M^2_h}{s-M^2_h}
\nn\\
&
+\kappa_V
\frac{4M^2_h}{\sqrt{(s-4M^2_Z)(s-4M^2_h)}}
\ln\l(\frac{s-2M^2_h-\sqrt{(s-4M^2_Z)(s-4M^2_h)}}
{2M^2_h\sqrt{1-\frac{4M^2_Z}{M^2_h}+\frac{sM^2_Z}{M^4_h}}}\r)
\biggr]
\,,\\
a_0&(h\pi^3 \to h\pi^3 )
\,=\,
\frac{s}{32\pi v^2}(\kappa_{VV}-\kappa^2_V)
\l[1-\frac{(M_h+M_Z)^2}{s}\r]
\l[1-\frac{(M_h-M_Z)^2}{s}\r]
\nn\\
&
-\frac{M^2_h}{16\pi v^2}\kappa_V
\biggl[
3\kappa_3-2\kappa_V
+\kappa_V\frac{M^2_h}{s-M^2_Z}
\nn\\
&
-\frac{3\kappa_3M^2_h s}{[s-(M_h+M_Z)^2][s-(M_h-M_Z)^2]}
\ln\l(1+\frac{[s-(M_h+M_Z)^2][s-(M_h-M_Z)^2]}{sM^2_h}\r)
\nn\\
&
-\frac{\kappa_VM^2_h s}{[s-(M_h+M_Z)^2][s-(M_h-M_Z)^2]}
\ln\l(\frac{s(s-2M^2_h-M^2_Z)}{sM^2_Z-(M^2_h-M^2_Z)^2}\r)
\biggr]
\,.
\end{align}

If we take $\kappa_V=\kappa_{VV}=\kappa_3=\kappa_4=1$ and $M^2_{W,Z}=0$, 
we obtain Eqs.~(3.1a)-(3.1g) in Ref.~\cite{Lee:1977eg}.

\paragraph{$T$-matrix}
The $S$-wave scattering matrix is calculated as
\begin{align}
\mathcal{T}_0
\,=\,
\l(
\begin{array}{cccc}
t_0(\pi^+\pi^- \to \pi^+\pi^-) & 
t_0(\pi^+\pi^- \to \pi^3\pi^3) &
t_0(\pi^+\pi^- \to h h) &
0
\\
t_0(\pi^3\pi^3 \to \pi^+\pi^-) & 
t_0(\pi^3\pi^3 \to \pi^3\pi^3) &
t_0(\pi^3\pi^3 \to h h) &
0
\\
t_0(hh \to \pi^+\pi^-) & 
t_0(hh \to \pi^3\pi^3) &
t_0(hh \to hh) &
0
\\
0 & 
0 &
0 &
t_0(h\pi^3 \to h\pi^3)
\\
\end{array}
\r)\,,
\end{align}
where
\begin{align}
&
t_0(\pi^+\pi^- \to \pi^+\pi^-)
\,=\,
a_0(\pi^+\pi^- \to \pi^+\pi^-)
\,\\
&
t_0(\pi^+\pi^- \to \pi^3\pi^3)
\,=\,
\frac{1}{\sqrt{2}}\,
a_0(\pi^+\pi^- \to \pi^3\pi^3)
\,,\\
&
t_0(\pi^+\pi^- \to hh)
\,=\,
\frac{1}{\sqrt{2}}\,
a_0(\pi^+\pi^- \to hh)
\,,\\
&
t_0(\pi^3\pi^3 \to \pi^3\pi^3)
\,=\,
\frac{1}{{2}}\,
a_0(\pi^3\pi^3 \to \pi^3\pi^3)
\,,\\
&
t_0(\pi^3\pi^3 \to hh)
\,=\,
\frac{1}{{2}}\,
a_0(\pi^3\pi^3 \to hh)
\,,\\
&
t_0(h\pi^3 \to h\pi^3)
\,=\,
a_0(h\pi^3 \to h\pi^3)
\,,\\
\nn\\
&
t_0(\pi^3 \pi^3 \to \pi^+\pi^-)
\,=\,
t_0(\pi^+ \pi^- \to \pi^3\pi^3)
\,,\\
&
t_0(hh \to \pi^+\pi^-)
\,=\,
t_0(\pi^+ \pi^- \to hh)
\,,\\
&
t_0(hh \to \pi^3\pi^3)
\,=\,
t_0(\pi^3 \pi^3 \to hh)
\,.
\end{align}

\bibliography{ref} 
\bibliographystyle{JHEP}
\end{document}